\documentclass[11pt,a4paper,english]{article}

\usepackage{graphicx}
\usepackage{authblk}

\usepackage{subfig}
\usepackage{morefloats}
\usepackage{multirow}

\usepackage{stackengine}
\newcommand{\Dbar}{\stackinset{l}{0.1ex}{c}{}{\rule{0.33em}{0.3pt}}{D}}

\begin{document}

\title{From Past to Future: Digital Methods Towards Artefact Analysis}
\date{}

\author[1]{Andrew Harris}
\author[2]{Andrea Cremaschi}
\author[3]{Tse Siang Lim}
\author[1,2]{Maria De Iorio}
\author[4]{Kwa Chong Guan}

\affil[1]{Department of Paediatrics, Yong Loo Lin School of Medicine, National University of Singapore}
\affil[2]{Singapore Institute for Clinical Sciences, A$^*$STAR}
\affil[3]{Independent Scholar}
\affil[4]{S. Rajaratnam School of International Studies, Nanyang Technological University}

\maketitle

\begin{abstract}
Over the past two decades, Digital Humanities has transformed the landscape of humanities and social sciences, enabling advanced computational analysis and interpretation of extensive datasets. Notably, recent initiatives in Southeast Asia, particularly in Singapore, focus on categorising and archiving historical data such as artwork, literature and, most notably archaeological artefacts. This study illustrates the profound potential of Digital Humanities through the application of statistical methods on two distinct artefact datasets. Specifically, we present the results of an automated die study on mid-1st millennium AD "Rising Sun" coinage from mainland Southeast Asia, while subsequently utilising unsupervised statistical methods on 2D images of 13\textsuperscript{th}-14\textsuperscript{th} century earthenware ceramics excavated from the precolonial St. Andrew's Cathedral site in central Singapore. This research offers a comparative assessment showcasing the transformative impact of statistics-based approaches on the interpretation and analysis of diverse archaeological materials and within Digital Humanities overall.
\end{abstract}

\section{Introduction}

The past two decades have seen an unprecedented data-driven revolution within the study of humanities and social sciences through computer-based analysis and the application of digital technology\cite{Berry_2011, Schriebman_etal_2004}. Known collectively as "Digital Humanities", this field is broadly defined by Drucker as encompassing "work at the intersection of computational methodology and humanities materials", and has been developed to better address humanistic questions, approaches, and data analysis \cite{Drucker_2021}. Originally known as 'humanities computing', Digital Humanities began as a series of early initiatives of data digitisation through the creation of textual archives and databases, but has evolved to encompass methods of data interpretation such as "computer-based statistical analysis, search and retrieval, topic modelling, and data visualisation" \cite{Berry_2019}. Given the recent advances in statistics and computer sciences, new tools are constantly being developed, each groundbreaking in the quantity of data that can be analysed and the quality of analysis that is performed (alongside efficiency and cost-effectiveness) that could otherwise not be achieved through manual, human study \cite{Mara_2022, Karl_etal_2022, Eslami_etal_2020}. Digital methods can be developed gradually and build in complexity, and can be supervised or automated, allowing scholars to focus their time and effort instead towards the creation of novel methodologies and interpretation of their data rather than exclusively data analysis. Digital Humanities also allows for substantial collections of data to be assessed through simplified workflows, for instance through Drucker’s model which streamlines data into digitisation (Materials), computational analysis (Processing), and publication (Presentation), striking a balance between ease of use and critical understanding \cite{Drucker_2021}. Finally, the demand for publicly  available data has led to methods which emphasise open interpretation and accessibility, creating “interpretive materials, the curation and documentation of objects, and the examination of the digital reception of heritage, particularly through social media”. \cite{Morgan_2022}

The use of digital methods within the study of humanities is readily applied to the various sub-fields that make up archaeology, defined as "the study of the ancient and recent human past through material remains" \cite{SAA}. Digital archaeology thus describes methods and theory that stem from the application of computational methods to archaeological studies, with the term “computational archaeology” readily applied to computer-based systems that “adapt existing technological innovations for specific archaeological purposes” \cite{Aycock_2021, Grosman_2016, Morgan_2022}. Mara \cite{Mara_2022} notes that the use of mathematical and statistical methods in archaeology occurred quite early, at the forefront of Digital Humanities \cite{Binford_1965, Clarke_1973}. Since the 1970s, digital methods have been used in a variety of archaeological investigations, for instance mapping archaeological sites through the application of GIS, the reconstruction of ancient sites and artefacts through 2D or 3D modelling, sorting and typologising various artefacts using unsupervised or semi-supervised statistical models, and the digitisation and archiving of museum and/or private collections \cite{Khunti_2018, Natarajan_etal_2023, Papaioannou_etal_2002, Rasheed_and_Nordin_2015, Tuno_etal_2022}.

The past decade has seen the rise of Digital Humanities in Southeast Asian contexts, most notably in Singapore, through the categorisation and archiving of historical datasets such as artwork, literature, artefacts, and archival texts\cite{vanLit_and_Morris_2024}. Many of these initiatives currently comprise the simple storage of data or the creation of public or private databases, but in select cases involve the use of digital applications in historical interpretation. Heng, for example, utilises keyword searches within ancient Chinese texts as a means for “data-trawling” specific topics related to Southeast Asia’s premodern history, for instance trends in trade between Southeast Asia and China during specific dynastic periods through referencing specific trade goods \cite{Heng_2019}. Moreover, Klassen et al. utilise linear-regression algorithms to predict the construction dates of temples in 9\textsuperscript{th}-14\textsuperscript{th} century Angkorian Cambodia from an incomplete set of foundation inscriptions \cite{Klassen_etal_2018}. Computational analysis, however, has never been applied to Southeast Asian artefacts recovered from archaeological contexts, which provides a unique avenue of inquiry and the application of novel statistical methods for analysis and interpretation.

To highlight the potential of digital humanities in Southeast Asian archaeological studies, this paper focuses on two separate datasets of artefacts from first and early-second millennium AD, both thought to represent patterns of trade and exchange across Southeast Asia’s “Silk Road of the Sea” \cite{Miksic_2013}. The first is a series of silver coins minted in the mid-1\textsuperscript{st} millennium AD featuring a uniform but locally varied “Rising Sun” motif. Although this motif likely originated in the ancient Pyu city-states of north-central Myanmar, Rising Sun coins have been excavated across mainland and peninsular Thailand and in select regions of Cambodia and Vietnam \cite{Epinal_2013, Epinal_and_Gardère_2014, Gutman_1978, Wicks_1992}. Using coins from the Konlah Lan hoard from the site of Angkor Borei in Takeo Province, Cambodia and coins from various excavations at the important ancient port of Oc Eo, An Giang Province, Vietnam, we present the results of an automated die study employing recently-proposed high dimensional clustering methods to assess whether coins found in specific and regional contexts may have been minted from the same dies, as well as to better understand the role of standardised coinage in this relatively under-explored ancient theatre of trade. 

The second dataset comprises of rim- and base-sherds from earthenware ceramic vessels excavated from the site of St. Andrew’s Cathedral in central Singapore \cite{Lim_2012, Miksic_2013}. These sherds come from the 14\textsuperscript{th} century settlement known in historical sources as Temasek, which for a short period acted as a notable locus of international maritime trade between China and India prior to the foundation of modern Singapore in the early 19\textsuperscript{th} century.  We show that unsupervised classification methods using 2D images can accurately recover group of sherds belonging to the same original vessel. 

The paper is structured as follows.

\section{Die Studies and Historical Background}
\subsection{Die Studies and Digital Numismatics}

Die studies are essential within the study of ancient numismatics, and are an important tool for assessing the economic history of early states \cite{Heinecke_etal_2021}. Ancient coins were typically minted from hand-engraved dies, with obverse-face (front) and reverse-face (back) moulds struck together over heated metal. Each face features unique, symbolic designs representative of familiar politico-religious symbols, or denoting a ruler themselves. These images suggest that, alongside an understood value, the initial conception and production of many currencies often reflect a specific identity ascribed to centralised governance \cite{Schaps_2015, Graeber_2011}. Die studies have the potential of quantifying the relative number of coins minted from any set of obverse or reverse dies, as well as mapping the production, distribution, and spread of coinage across geographic areas \cite{Esty_2011, DeCallatay_1995}.

From a technical point of view, die identification can be seen as a classification problem with a finer subdivision of classes; for instance, every coin class consists of several different dies used for striking. The appearance of an ancient coin is usually unique because no two blank coins were ever struck at exactly the same angle or with the same force and dies deteriorated over the course of the minting process, resulting in different impressions. Other factors influencing variation in coin appearance include material, craftsmanship, tools used in minting, the specific die, mint signs and shape. Furthermore, coins suffered from wear and tear, were clipped to shave off precious metal, or were marked by money changers and government authorities. As a result, specimens struck from the same die may show a large degree of variability, whereas coins of the same type may look very similar, even when struck from different dies, because all dies of the same issue were based on a common prototype. Apart from these factors, comprehensive die studies themselves are an enormous time investment, and depending on the number of coins involved, are estimated to take between a year (1,000+) to a lifetime (60,000+ coins) \cite{Aagard_and_Marcher_2015, Van_Alfen_2017}. Furthermore, any individual die study represents a regional subset of any total output, and requires contextualisation within a greater body of known examples.

Digital numismatics is an approach to the study and appreciation of coins and currency exploiting computational techniques to enhance our understanding and engagement with numismatic artefacts. This field, that has received increasing attention over the last decades, combines the traditional expertise of numismatists with cutting-edge digital tools and techniques, opening new avenues for research, cataloguing, and accessibility. One of the key strengths of digital numismatics is its ability to create comprehensive and easily accessible databases of coins from various historical periods and regions. This not only aids scholars and collectors in their research endeavours, but also democratises access to this rich cultural heritage for a broader audience. Advanced imaging technologies, such as high-resolution photography and 3D scanning, are used for detailed and accurate documentation of coins, revealing intricate details that might otherwise be overlooked \cite{Bentkowska_MacDonald_2018, Hess_etal_2018}. This not only aids in characterisation and authentication but also provides a valuable resource for the study of iconography, inscriptions, and historical context. 

Heinecke et al. emphasise that automated die studies, as opposed to manual examinations (even with photographs) would both help categorise far more efficiently, and possibly more accurately, than those completed with the human eye, even with supplemental verification by the appropriate numismatic experts \cite{Heinecke_etal_2021}. In recent years, computer vision-based analysis of ancient coins has been attracting increasing attention, yet despite this research effort the results achieved remain poor and far from being useful for any practical purpose \cite{Cooper_and_Arandjelovic_2019}. The application of computer vision methods to coin analysis has mainly focused on coin classification \cite{vanderMaaten_and_Postma_2006, Zaharieva_etal_2007, Kampel_and_Zaharieva_2008, Anwar_etal_2015, Sasi_and_Sreekumar_2015, Aslan_etal_2020}. Moreover, the existing methods usually rely on the availability of a reference set, a tall order for pre-modern coinage, which comes in hundreds of thousands of distinct types. Most success has been obtained with contemporary machine-made coins (e.g., the Dagobert coin recognition system aimed at sorting high volumes of coins \cite{Nölle_etal_2003}), but algorithms for the classification of modern coins are not directly applicable to hand-minted pre-modern coins. The assumption of uniformity, and thus the straightforward feature comparison for modern coins, facilitates the classification process substantially. Most existing algorithmic approaches to coin identification are based on the extraction of local features from each coin image \cite{Cooper_and_Arandjelovic_2020} and on the definition of a similarity measure between pairs of coins, which is crucial for classification accuracy. These approaches often result in poor performance due to loss of spatial relationships between the different impressions on the coin. To the best of our knowledge, only a limited number of the existing proposals explicitly address the problem of die analysis, i.e. the determination of the number of dies corresponding to a sample of coins and their partition according to dies \cite{Natarajan_etal_2023}. Moreover, current methods are mainly based on pairwise comparisons and at best use heuristics to determine the number of dies \cite{Taylor_2020}. Finally, there has thus far been no attempt to statistically estimate chronology by combining reverse and obverse information.
 
\subsection{A Brief History of 1\textsuperscript{st} Millennium AD Southeast Asian "Rising Sun" Coinage}
We  demonstrate the efficacy and potential of automated die studies using an unsupervised statistical model on a sample of coinage from 1\textsuperscript{st}millennium AD Southeast Asia, a region featuring a relatively under-explored numismatic history compared to that of contemporary currency-based economies such as Rome, India, and Central Asia \cite{Cribb_and_Bracey_2019, Sutherland_and_Carson_1926-2019, Mitchiner_2002}. Archaeologist John Miksic, among others, \cite{Miksic_2013, Carter_etal_2021} notes that Southeast Asia formed an important midpoint for international maritime trade between India and China as early as the 2nd millennium BC, while Roman and Chinese sources written as early as the 3rd century BC document Southeast Asia’s trade routes, ports, goods, tribute, and to a lesser extent ancient peoples and customs active in the region \cite{Vickery_2003, Heng_2019}. Miksic points out that Southeast Asians were evidently important facilitators of trade during this period \cite{Miksic_2013}, yet Mahlo  argues that “[scholars] are still dealing largely with peoples, kingdoms, and ruling dynasties with unknown histories” \cite{Mahlo_2012} and thus the internal workings and dynamics of trade are often unknown or at least require further investigation through archaeological analysis.

Southeast Asia’s earliest locally-produced coinage, denominations of 98\% pure silver, was first minted in the Pyu city-states of southern and western Myanmar as well as the contemporary Kingdom of Arakan in northeastern coastal Myanmar as early as the 4\textsuperscript{th} century AD \cite{Gutman_1976, Gutman_1978}. The introduction of coinage to Southeast Asia is not believed to have been a local innovation, instead a product of the gradual process of Indianisation across Southeast Asian, which through trade and cultural exchange resulted in the integrated South Asian religion with indigenous Southeast Asian systems of local rulership and spirit-worship \cite{Coedes_1968, Mus_1933}. With the exception of Arakanese coins,  Southeast Asian coinage from this period was not inscribed with the names or images of rulers. Instead, coinage featured a standardised set of symbolic Hindu-Buddhist images; while slightly augmented over time, these remained fairly uniform throughout their period of production. For instance, what are thought to be the earliest Southeast Asian coins feature the image of an auspicious conch \textit{(sankha)} (obverse) and an aniconic Indian symbol known as a \textit{srivatsa} (reverse), thought by Gutman to represent the Hindu goddess Sri \cite{Gutman_1978, Goyal_1995, Wicks_1992, Miksic_and_Goh_2017}. 

The most widespread type of coin produced during this period, and the focus of this study, depicts the Rising Sun (obverse) and \textit{srivatsa} again on the reverse; the former features a half-risen sun surrounded by a border of 27-31 pellets, while the \textit{srivatsa} is flanked by a swastika (a Hindu-Buddhist symbol of good fortune) and \textit{bhadrapittha} (drum of Shiva) and situated below a small sun and moon of varying design (Figure 1) \cite{Mahlo_2012, Liebert_1976}. “Rising Sun” coins, measuring between 27-32mm (known avg. 29.75g) and weighing between 59-13.3g (known avg. 8.975g), are believed to have been predominantly manufactured and distributed between the 4\textsuperscript{th}-7\textsuperscript{th} centuries AD. These coins are not believed to represent any single Pyu ruling family nor monarch of this period but instead symbolise rulership in general; recounting this period, the Burmese Glass Palace Chronicle of 1796 AD refers to the separate Pyu kingdoms collectively as the “Sun Dynasty” \cite{Gutman_1978, Tin_and_Luce_1976}.

Rising Sun coins of a variety of subtypes have been found across mainland Southeast Asia from eastern Bangladesh to southern Vietnam \cite{Beaujard_2019}; however, regional groupings and relative chronologies have been established only on the basis of coin metrics (weights/measures) and the iconographic programme found on the coins themselves \cite{Gutman_1976, Gutman_1978, Wicks_1985}. Often, the find-sites of these coins are within settlements verified as important regional economic centers, and are either inland riverine settlements or coastal ports \cite{Hall_1999}. Excavations at sites such as Beikthano, Halin, and Sri Ksetra (Myanmar), U Thong, Pathom, and Lopburi (Thailand), and Oc Eo (Vietnam, see below) report numerous findings of Rising Sun coins; however, archaeological finds in Southeast Asia have not been widely reported \cite{Mahlo_2012, Onwimol_2018}. Nevertheless, when publicised, coin finds are either noted as single examples, often found surrounding important religious monuments, or as part of buried hoards, conglomerations of coins deposited in ceramic vessels or other containers\cite{Wicks_1992, Epinal_2013}. Hoards, which in Southeast Asia are known to number between 3-2000+, are incredibly important for further assessing artistic typologies beyond existing private collections \cite{Mitchiner_1998, Mitchiner_2002, Mahlo_2012}, as well as for determining relative minting dates, whether coinage was produced locally or imported, and even how wealth and power were perceived vis-à-vis coinage in various regions, for example through the production of smaller denominations vs. the fractioning of coins as bullion \cite{Mitchiner_2002}. 

The coins analysed within this study represent the easternmost known extent of  “Rising Sun” coin finds in Mainland Southeast Asia, which in turn mark the easternmost reach of any polities which used these coins as either currency or weighted standard of silver for trade. As noted, each of the samples assessed in this study was found in the area of the Mekong Delta at two sites: the inland centre of Angkor Borei in Takeo Province, Cambodia, and within and around the ancient port-city of Oc Eo in An Giang Province just north of the Mekong Delta. Both sites were inhabited contemporaneously as settlements within the Kingdom of Funan (c. 2\textsuperscript{nd}-7\textsuperscript{th} centuries AD), an important early Indianised polity in Southeast Asia documented in contemporary Chinese chronicles \cite{Malleret_1959-1963, Stark_2004, Vickery_2003}. Funan is thought to have encircled the Mekong Delta and Gulf of Thailand and provided strategic links between Indian and Chinese spheres of coastal trade \cite{Coedes_1968, Heng_2019}. However, despite residing in the same cultural sphere, connected by an artificial canal\cite{Sanderson_etal_2007}, the archaeological contexts within which these coins were found differ greatly. For instance, the majority of known Rising Sun coins from Angkor Borei come from a single hoard found in 2011 at the find-site of Konlah Lan near the ancient settlement’s western gateway \cite{Epinal_and_Gardère_2014}. Originally numbering over 2000 in total according to local reports, at the time of writing the Konlah Lan coin hoard represents arguably the most significant  early Southeast Asian coinage finds ever recorded; however, less than 100 full coins alongside several fractional denominations are now housed in Cambodia’s SOSORO Museum of Economy and Money in Phnom Penh \cite{Epinal_and_Gardère_2014}. The coins from Oc Eo and in the surrounding countryside of An Giang Province, meanwhile, were found during multiple excavations directed by French archaeologist Louis Malleret in the 1940s \cite{Malleret_1959-1963}. Unlike the Konlah Lan hoard and other finds around Angkor Borei, which primarily comprised of full-unit denominations, Oc Eo's excavations unearthed proportionately far more cut coin fragments (1/2, 1/4th, 1/8th, and 1/16th); 251 of the 262 coins samples housed at the Ho Chi Minh City (HCMC) History Museum from Oc Eo were fragmented compared to only twenty-four recovered from the inland site \cite{Epinal_and_Gardère_2014}. Consequently, only 12 full Rising Sun coins from Oc Eo/An Giang have been incorporated within this study, and collectively represent the total documented from this archaeological region \footnote{Two others have been recovered from an undocumented site outside of Ho Chi Minh City (Saigon) in 1886, although their current whereabouts are unknown \cite{Cappon_1886}}.


\subsection{Data}

Data collection for this study was completed between November 14\textsuperscript{th}-17\textsuperscript{th}, 2022 at the SOSORO Museum of Economy and Money in Phnom Penh, Cambodia and February 21\textsuperscript{st}-24th\textsuperscript{th}, 2023 at the Ho Chi Minh Museum of History, Vietnam. Coins were measured for diameter/radius and thickness, weighed, and photographed. A test-set of 90 full-denomination Rising Sun coins (88 obverse, 86 reverse)\footnote{Due to issues with accessing the Oc Eo Rising Sun coins, only one face from six of the coins (4 obverse, 2 reverse) are featured in the study.} was included into the study, with 78 coins from Angkor Borei, presumably from the Konlah Lan hoard, and 12 from Oc Eo/An Giang Province.  Each coin was given a classification number beginning with RS\footnote{"Rising Sun", abbreviated.}, ex. RS0100.

A comparative manual die study of the 90 coins within this test-set was completed  prior to the automatic study and used as benchmark. All coins were compared with examples from private collections (\cite{Mahlo_2012, Mitchiner_1998, Mitchiner_2002} and online auction catalogues; the latter have been consolidated through the Southeast Asian Numismatics Digital Archive by renowned Southeast Asian numismatic expert Robert Wicks \cite{www.seanda.omeka.net}. A larger study incorporating coins from museums in Myanmar, Thailand, and the United Kingdom is ongoing.

\subsection{Pre-processing}\label{sec:Coins_preprocessing}

\begin{figure}[ht]
	\centering
	\subfloat[Obverse]{\includegraphics[width=0.5\textwidth]{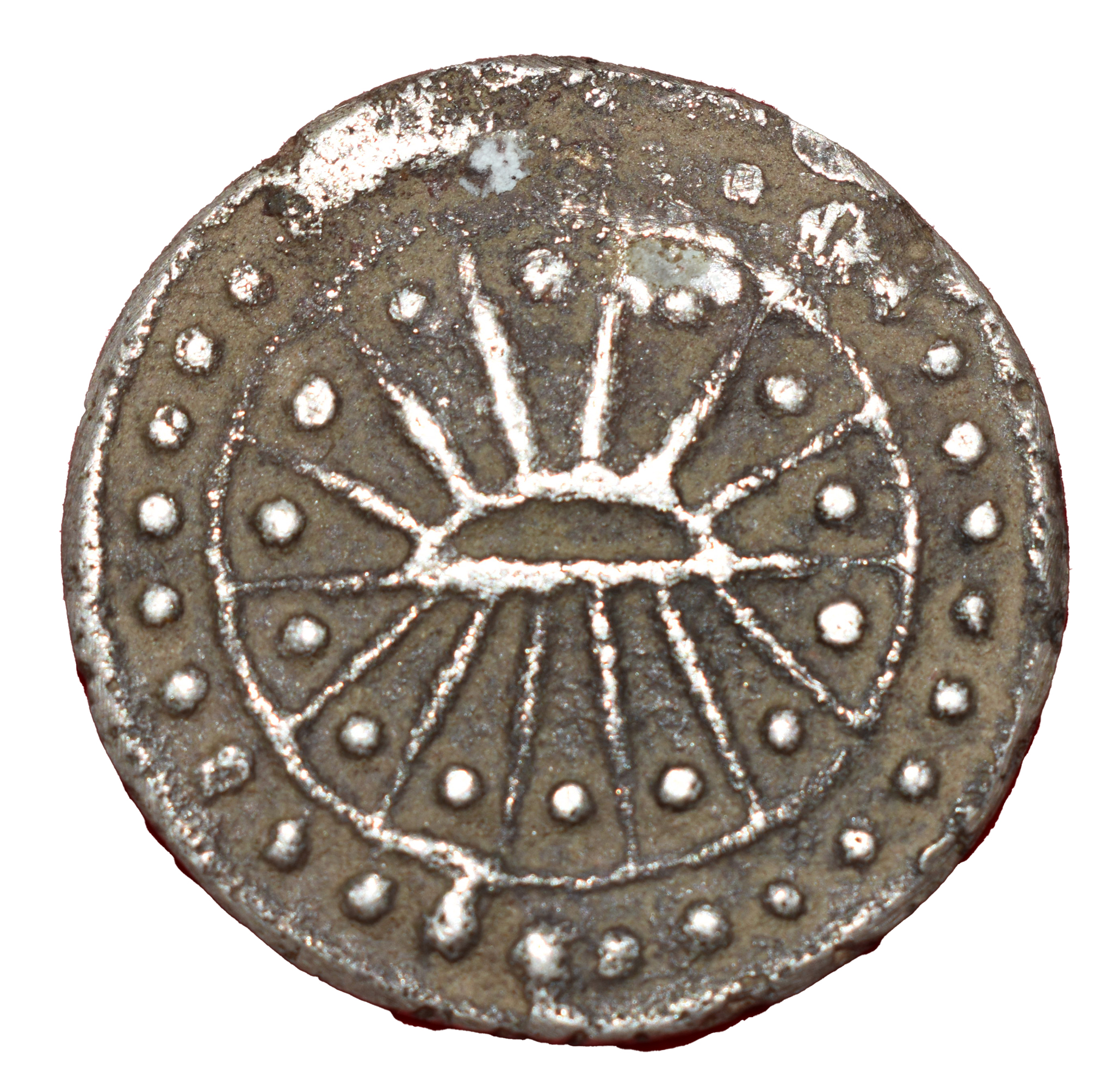}}
	\subfloat[Reverse]{\includegraphics[width=0.5\textwidth]{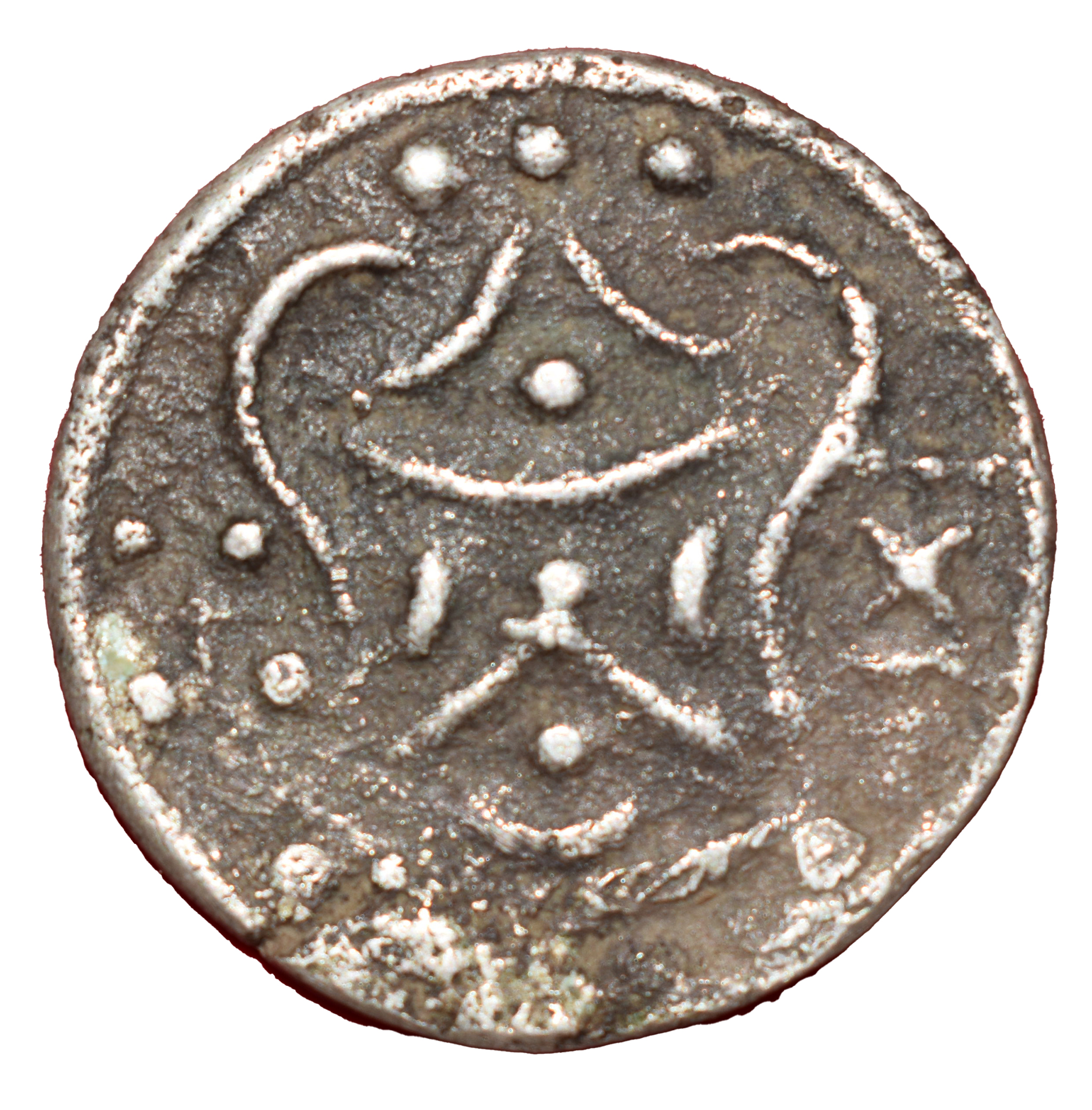}}

    \subfloat[Obverse]{\includegraphics[width=0.5\textwidth]{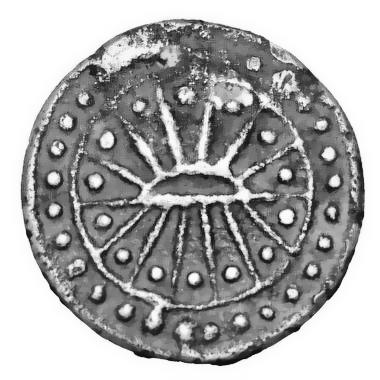}}
	\subfloat[Reverse]{\includegraphics[width=0.5\textwidth]{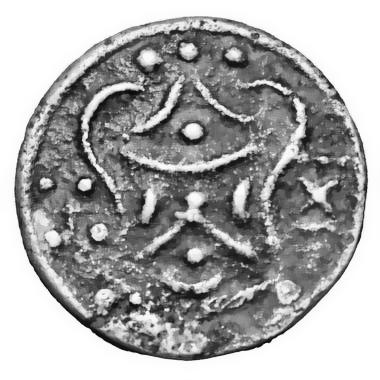}}
	\caption{Example of a coin  before (top row) and after (bottom row) pre-processing enhancement. Each row shows the obverse and reverse side of the coin.}
	\label{fig:example_coin}
\end{figure}

We performed a series of pre-processing steps following Natarajan et al., 2023 \cite{Natarajan_etal_2023}. Specifically, we convert the images to greyscale and resize them to $300 \times 300$ pixels. Then, we apply total-variation image restoration (Rudin et al. , 1992 \cite{Rudin_etal_1992}) to each image to reduce the noise introduced during image acquisition resulting from low lighting, and to even out small aberrations, while preserving the quality of edges and corners in the images. Contrast limited adaptive histogram equalisation (Pizer et al., 1987 \cite{Pizer_etal_1987}) is then applied to locally enhance contrast and edge definition, while limiting noise amplification in near-constant image regions. Potential artefacts introduced by the local transformations used in the latter procedure are then reduced by a second application of total-variation restoration. See Figure \ref{fig:example_coin} for an example of raw versus enhanced coin images (obverse and reverse).

\subsection{Statistical methodology}

In this work, we exploit techniques from machine learning and Bayesian statistics to extract coin features and quantify differences between pairs of coins. This approach allows the specification of suitable distances to be used in distance-based clustering, with the aim of grouping the coins based on their features. In particular, we employ the D2-Net model proposed by Dusmanu et al. 2019 \cite{Dusmanu_2019a, Dusmanu_2019b}, performing dense feature extraction via a previously trained Convolutional Neural Network, namely the 16-layers network VGG16 proposed by Simonyan and Zisserman, 2014 \cite{Simonyan_and_Zisserman}. D2-Net yields a representation of the coin image that is simultaneously a detector (i.e., a locator of keypoints in the image) and a descriptor (i.e., represented by an array of features describing the image properties at the corresponding keypoints). More details, as well as a GitHub repository with code and guidelines for implementation, can be found in Dusmanu et al., 2019a \cite{Dusmanu_2019a} and Dusmanu et al., 2019b \cite{Dusmanu_2019b}.

\begin{figure}[ht]
	\centering
	\includegraphics[width=1\textwidth]{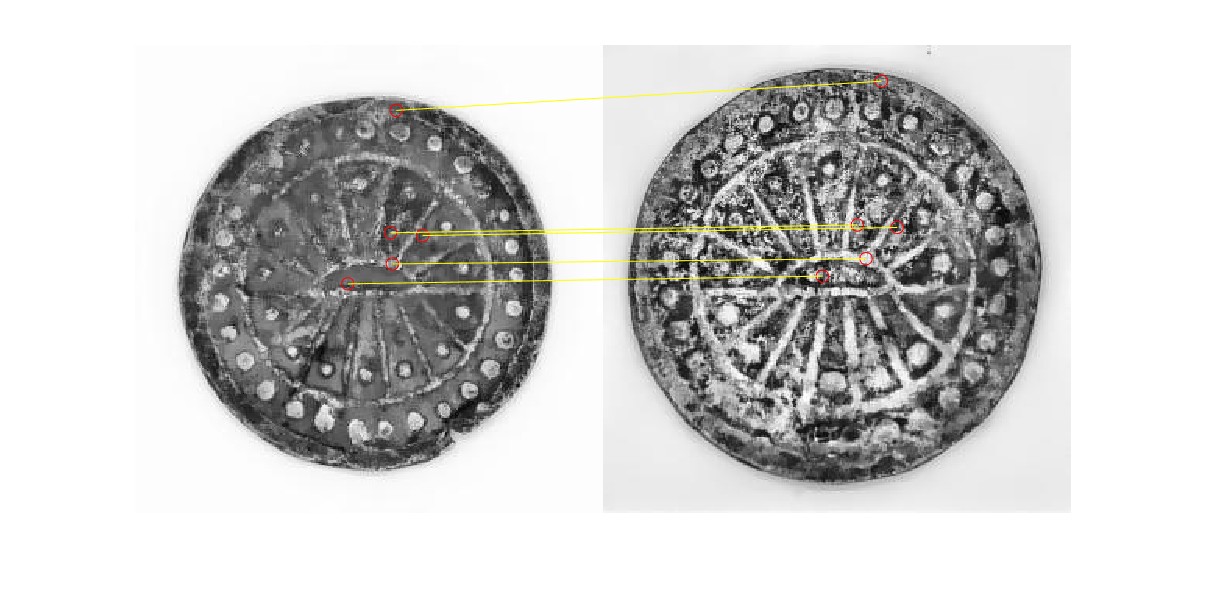}\\
    \includegraphics[width=1\textwidth]{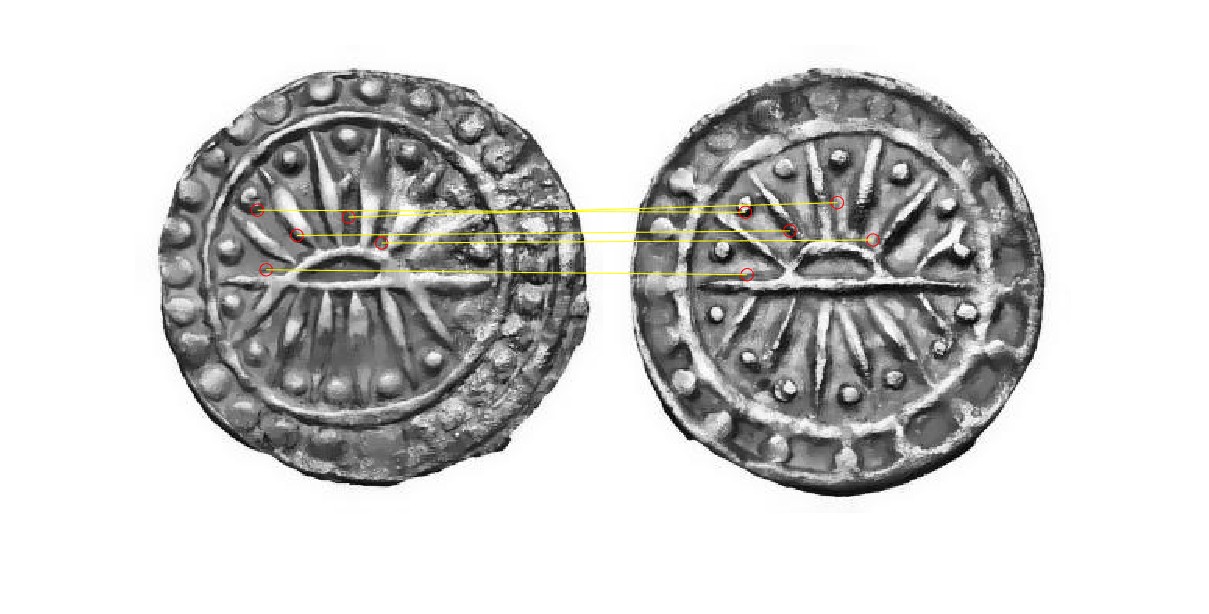}
	\caption{Example of landmark matching between two coin obverses belonging to the same (top) or different (bottom) die set.}
	\label{fig:example_coins_landmarks}
\end{figure}

After feature extraction, we match coin image pairs via an efficient approximate nearest neighbour search \cite{Muja_and_Lowe_2009}. This results, for each image pair $(i,j)$, in a subset of matched \textit{landmarks}, selected among the D2-Net keypoints, between images $i$ and $j$. 
We then  identify the circular regions of each pair of images $i$ and $j$ enveloping the coins via the \texttt{Matlab} function \texttt{imfindcircles}. Unmatched keypoints and those located outside the coin circles, are discarded. An example of the resulting landmarks identified for a pair of coins (obverse) is shown in Figure \ref{fig:example_coins_landmarks}. 
The landmark features are used to compute different dissimilarity measures between pairs of coin images. First, the matched landmarks are ranked using Gaussian Processes landmarking \cite{Gao_etal_2019a, Gao_etal_2019b}, to identify the most reliable landmarks for image comparison. Thus, the features extracted by the D2-Net model corresponding to the ranked landmarks are used to compute a variety of similarity  metrics between coins. These include the number of matched landmarks, weighted Euclidean distances \cite{Natarajan_etal_2023}, as well as the pairwise structural similarity index measure (SSIM), comparing the degree of similarity between two images by weighing their luminance, contrast and structure \cite{Wang_etal_2004}. Of all the possible pairs of coins to be compared, we exclude those with less than three matched landmarks. The order of the landmarks matters in the computation of some of these measures, which include rank-dependent weights. The metrics are then used to compute a final dissimilarity measure corresponding to the Euclidean norm of the vector whose components are the individual metrics. This yields a dissimilarity matrix, composed of the pairwise distances between coin pairs. Due to the exclusion criterion based on the number of matched landmarks, the dissimilarity matrix presents some missing entries, which are imputed following the ultrametric procedure \cite{Makarenkov_and_Lapointe_2004}.

The dissimilarity measure is employed in the distance-based clustering approach proposed by Natarajan et al., 2023 \cite{Natarajan_etal_2023}. In their work, the authors propose a Bayesian nonparametric model for distance-based clustering where a suitable likelihood is specified for the pairwise distances between high-dimensional objects, thus drastically reducing the curse of dimensionality. Moreover, the prior distribution on the partition has the micro-clustering property \cite{Betancourt_etal_2022}, which allows the identification of smaller, but still informative clusters. In this context, a cluster of coins represents coins minted from the same die.  

The distance-based clustering model \cite{Natarajan_etal_2023} is fit to obverse and reverse images, independently and inference is performed through a Markov chain  Monte Carlo (MCMC) algorithm. See \cite{Natarajan_etal_2023} for details. 
The obverse and reverse MCMC chains are used to obtain posterior estimates of the partition of the coins into clusters by minimising the variation of information loss function \cite{Meila_2007}, which combines information within each of the two partitions (entropy), as well as the information shared between the partitions (cross-entropy). The estimated clustering of the obverse and reverse coin images are formed of four and three clusters of sizes (59, 14, 7, 5) and (73, 7, 5), respectively. To assess the level of accuracy of the proposed methodology, we compute the Rand index \cite{Rand_1971} between the estimated partitions and the  results from the manual analysis. Given two partitions of the same objects, the Rand index measures the proportion of pairs that are clustered in the same way (i.e., together or separately in both partitions) over the total number of possible pairs. This index takes values in the interval (0,1), with higher values indicating a higher agreement between the two partitions under comparison. We obtain Rand indices equal to 0.77 and 0.84 for obverse and reverse, respectively.

\subsection{Discussion}

\begin{figure}[ht]
	\centering
	\includegraphics[width=1\textwidth]{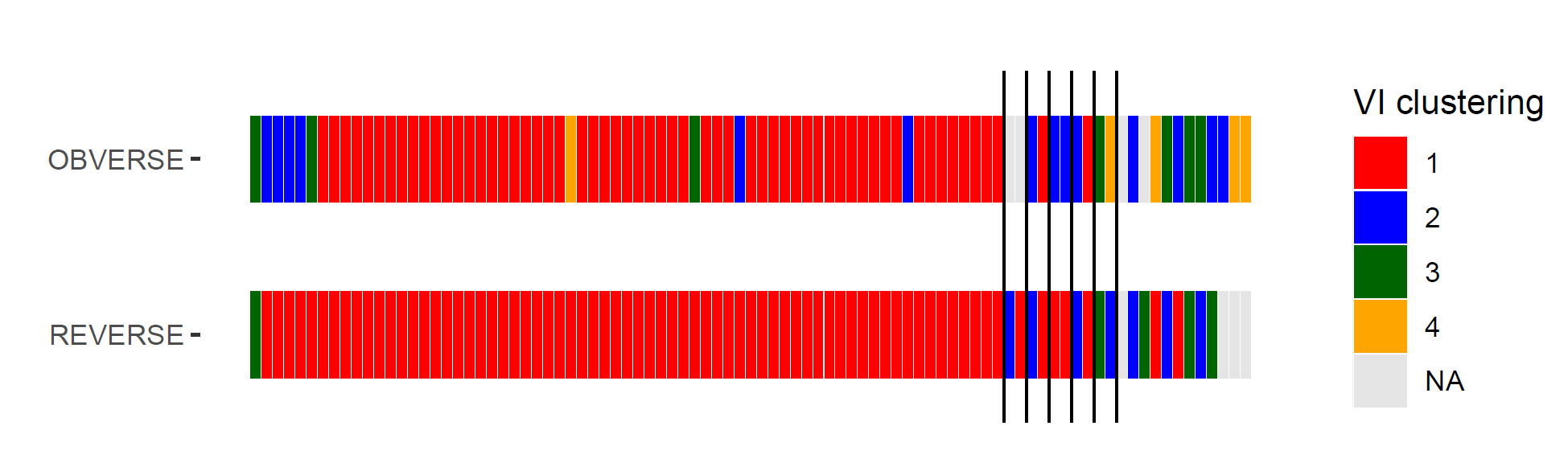}
	\caption{Clustering assignment obtained with the automatic analysis. Each element of the grid represents one of the 89 distinct coins analysed. The coins are displayed by obverse/reverse (top/bottom row) and each colour represent a different cluster. The black vertical lines separate coins belonging to different dies as by manual analysis, with the rightmost coins representing the singleton group. Grey grid elements indicate coins for which only one  face was available in the dataset.}
	\label{fig:Clustering_rows}
\end{figure}

From the manual die analysis,  we concluded that 67 coins sampled from the Konlah Lan hoard were minted from the same die, with an additional eight derived from four different obverse and reverse dies (with a pair of coins assigned to each die). The remaining three coins featured overlap between the obverse and reverse matches; coins numbered RS0212 and RS0760 in this study share an obverse die but not reverse, and coins RS0760 and RS0188 share a reverse but not obverse. This observation appears to be unique within studies of Rising Sun coins, as die assignment is almost always consistently the same between obverse and reverse sides. The twelve coins from Oc Eo, meanwhile, are entirely singletons. Finally, the result from the automatic data analysis are in general in line with those obtained by the manual analysis for both reverses and observes, in particular for the 67 coins coming from the same die. A summary of the comparison between the two analyses is displayed in Figure \ref{fig:Clustering_rows}. We notice that the automatic analysis is able to recover correctly the mix between the reverse and obverse discussed above.

Although automatic die analysis tools still need to be perfected, from a numismatic point of view, the availability of the automatic results allows for a faster sorting of coins in advance of a critical examination. The reduction in necessary comparisons results in saving numerous hours of manual labour. This is primarily because, on average, the remaining comparisons are less time-consuming than those required in a brute-force die study. A significant portion of images is already correctly assigned, and many accurate assignments have been made. In practical terms, visual validation mainly involves making corrections for low-quality images or poorly preserved coin images. In summary, automatic die analysis empowers researchers to dedicate more time to the numismatic and art-historical analysis of their material. Additionally, it provides valuable information about the minting process through die combination diagrams \cite{Heinecke_etal_2021, Esty_1986} and offers insights into mint output via statistical analysis. 

Historically, this study allows for examinations and scrutiny of previous typologies of Rising Sun coins as well as the significance of different coin types and subtypes at various archaeological sites in mainland Southeast Asia. For instance, comparisons with coins typologies generated from the private collections of Mitchiner \cite{Mitchiner_1998, Mitchiner_2002} and to a greater extent Mahlo \cite{Mahlo_2012} suggest that the variations of coins sorted from Angkor Borei and the Konlah Lan hoard collectively represent Type 8b in Mahlo's typology, while the singletons from both Angkor Borei and the Mekong Delta region in Clusters 2-4 are a combination of subtypes 8a, 8b, and 8c \cite{Mahlo_2012}. This typology is based on the downgrading of the technical quality and “errors” in the figuration from an “idealised” example (8a) minted in Myanmar rather than more detailed analysis of coins and their archaeological find-sites. The 67 coins of identical dies, for instance, suggest that these broad 8a, b, and c typologies can be broken down into more detailed, localised subtypes connected to both geography and regional activity. It is also probable that the 8a coins found from Oc Eo were imported, which suggests that the Funanese port played host to trade from Pyu city-states and beyond; coins found at Angkor Borei in the Konlah Lan hoard, meanwhile, were probably locally-minted imitations of a standardised silver bullion rather than as a centrally-issued currency, and were used for trade and to signify wealth through association with an early trans-regional Southeast Asian maritime economy \cite{Onwimol_2018, Carter_etal_2021, Moore_2009}. Evidence for the local production of early coinage is scarce in Cambodia, but an example of a stone-carved Rising Sun obverse mould is housed at Bangkok's Coin Museum, presumably found at a site in Thailand \cite{Onwimol_2018}, and the overlapping obverse and reverse designs between coins RS0188, RS0212, and RS0760 indicates that several series of these coins were indeed likely minted locally beyond the influence of Pyu centres \cite{Wicks_1992}. Finally, Epinal \cite{Epinal_and_Gardère_2014}  suggests that the Konlah Lan hoard may have been buried in a period of political instability during the 7\textsuperscript{th} century, one which, documented in Chinese sources, saw the eclipse of Funan and the rise of a polity or polities known as "Chenla" or "Zhenla". \cite{Coedes_1968, Vickery_1998}. Chenla's pre-eminence, similar to the later dominance of Burmese Pagan (849-1297 CE) and Cambodian Angkor (801-1431 CE), saw the establishment of insular, centralised agrarian states across Southeast Asia ruled by Hindu or Buddhist monarchs and the subsequent limited output and use of coinage in the Mekong Delta region as regional trade between these polities dwindled. Thus, although no dateable material has survived to verify the time-frame of deposit of the Konlah Lan hoard, Epinal's interpretation suggests the possibility that the 8b sub-variants, especially those 67 matching coins primarily from Cluster 1, were some of the last iterations of Rising Sun coins minted in 1st millennium AD Cambodia.

\clearpage

\section{Ceramics}

\subsection{Statistical Analysis of Ceramics}
Ceramics (pottery) are an artefact class defined as any shaped object made of fired clay, and represent arguably the most numerous artefacts collected in archaeological investigations. The most common archaeological ceramics are typically ceramic vessels, which are typically broken into fragments known as sherds that appear either as surface-finds surrounding areas of occupation or within excavation trenches in strata corresponding with occupation. Ceramics are typically fired at low-temperature (earthenware) or high-temperature (stoneware), and are sometimes made from specially-sourced clay (for example kaolin porcelain). The remains of vessels such as pots, dishes, and vases provide significant information about the daily lives of people in the past, and their study represent an important inquiry in archaeology; Karl notes that "pottery analysis in archaeology addresses many topics ranging from the resources of the potter’s clay, the forming of pottery, the vessel shapes and painting styles - including their development over time – to its use, trade, discard and reuse" \cite{Karl_etal_2022}. The sherds which survive from any of these items, however, are often in a poor state of preservation, either broken or worn-off beyond recognition, and archaeologists rarely find ceramics in an optimal state for direct visual examination. Furthermore, ancient potteries are almost always excavated with missing parts, and both analysing and reconstructing these vessels manually is time-consuming and often expensive \cite{Mara_2022}. 

Given the significance of ceramic study to the archaeological field, it is no surprise that many of the digital archaeological methods generated over the past two decades have been applied to reconstructing pottery objects from either sorted or arbitrary collections of excavated sherds\cite{Kampel_and_Sablatnig_2003a, Kampel_and_Sablatnig_2003b, Rasheed_and_Nordin_2015, Eslami_etal_2020, Eslami_etal_2021}. Papaioannou et al. were the first to propose methods for automatic three-dimensional (3D) construction based on the alignment of parts focusing on surface geometry, whereby matching was done directly through the broken edges between two arbitrary fragments utilising optimisation algorithms \cite{Papaioannou_etal_2002, Rasheed_and_Nordin_2015}. Eslami et al. note that the majority of subsequent digital ceramic reconstruction methods have relied on the procurement of these types of 3D images of sherds, either through the stretching of images or 3D modelling (for example through photogrammetry or laser scanned images) to incorporate all possible geometric features \cite{Eslami_etal_2020}. Kampel and Sablatnig, meanwhile, introduced an automated archival, classification, and reconstruction system using the front and back of single fragments, and automatically extracted the features of what was suggested to be the complete vessel by computing the profile in relation to the documented measurements and/or  the ratios of the dimensions of the object \cite{Kampel_and_Sablatnig_2003b, Karl_etal_2022}. 

Despite the time-saving and cost-effectiveness of 3D image innovations compared to manual ceramic sorting and reconstruction, Eslami et al. suggest further improvement, noting that 3D reconstruction methods form a “bottleneck in the automatic analysis of large quantities of fragments. From...direct experience..., the process to get a valid discrete geometric model requires, for each sherd, a mean time of about 30 minutes” \cite{Eslami_etal_2021}. Instead, Eslami et al. recommend a shift towards the use of 2D images of front profiles and edge curves. This method was previously utilised by Richter et al. for reconstructing torn documents to recover texts, as well as by Brown et al. for restoring prehistorical frescoes found in excavations at the site of Thera, Greece where information such as shape, colour, texture, and roughness was recorded and utilised alongside photographs and measurements \cite{Brown_etal_2008, Richter_etal_2011}. As such, recent evaluations of the field of digital ceramic reconstruction \cite{Eslami_etal_2020} \cite{Karl_etal_2022} emphasise that advances in automated analysis through statistical methods might be more effective for 2D ceramic reconstruction. In our application in Section~\ref{sec:2Drec}, we followed the 2D approach.

\subsection{Historical Background}

The ceramics analysed in this study come from excavations conducted surrounding St. Andrew's Cathedral in central Singapore, an area of ancient settlement predating the foundation of modern Singapore in the early-19\textsuperscript{th} century. Over the last 40 years, large amounts of artefacts dating approximately between the 13\textsuperscript{th} and 14\textsuperscript{th} centuries AD have been recovered from several archaeological sites on Singapore Island \cite{Miksic_2013}).  These sites are all located within an area bound by the city-state's Fort Canning Hill to the north, the Singapore River to the west Stamford Road to the east and the Straits of Singapore to the south.  Coincidentally, Malay oral tradition and written records, most notably the Malay Annals (Sejarah Melayu (SM)), claim that a polity known as ‘Singapura’ was established on the island of ‘Temasek’ by a semi-divine prince around the same region and period.  From as early as 1819, these artefacts have been associated by historians and scholars alike with a hypothetical settlement and polity derived from inferences drawn from the SM about ‘Temasek-Singapura.’ It is the present consensus that these artefacts belong to a complex port-city which once existed on Singapore Island in precolonial times, which is validated by both the SM and the 14\textsuperscript{th} century \textit{Description of the Barbarians of the Isles} written by Chinese voyager Wang Dayuan in 1349 \cite{Rockhill_1914}. Despite almost two centuries of both historical and archaeological research however, very few of the political, cultural and socio-economic aspects of this settlement, its relationship with neighbouring polities and its significance within Southeast Asian precolonial history can be confirmed by verifiable historical evidence. In turn, despite the notable innovations in digital humanities seen in Singapore in recent years \cite{Heng_2019}, archaeology as an academic field currently remains underdeveloped despite its promising inception in the 1980s.

These ceramics were excavated from the St. Andrew's Cathedral site (STA) between 2003 and 2004, which at the time comprised the ‘most extensive, longest-lasting, as well as the most systematically designed and executed archaeological project in the history of Singapore’ \cite{Miksic_and_Lim_2003}. A team of archaeologists led by Professor John Miksic of the National University of Singapore initiated an archaeological project to recover any artefacts at that site before they were lost in the redevelopment of the site into a two-level basement extension of the cathedral in December 2003 (\cite{Pamphlet_2003}). Over a period of 28 weeks between September 2003 and June 2004, the team surveyed and excavated the site within three consecutive phases: augering, test-pit excavation and salvage excavation. The entire site occupied an area of approximately 2,400 square feet (223 m$^2$) – northwest of the Cathedral, along Singapore's North Bridge Road – and a total of 190 excavation units measuring 2x1m were completed by the end of the project and excavated either down to the point of artefect sterility (absence of artefacts) or the level of the water table (180cmbd\footnote{Centimeters below datum, an arbitrary xyz coordinate set by the excavators of any archaeological trench.}) \cite{Lim_2012}.

The scale of the project yielded proportionally significant results: over 330,000 pieces of artefacts from the precolonial to colonial periods, estimated to weigh around 1 ton (0.91 metric ton), were recovered from 1009 metric tons (or 636 m\textsuperscript{3}) of excavated soil. This amount was estimated to equal ‘the amount of artefacts recovered in all previous archaeological research in Singapore’ \cite{Miksic_and_Lim_2003} up to 2003. In other words, the STA site had an average artefact density of 519 artefacts per cubic metre; of this amount at least half can be ascribed to the Temasek-Singapura period.

It is clear from the stratigraphic soil profiles of each trench and artefact yields alongside historical sources that three phases of human occupation exist on the STA site: colonial (1819-1965), post-colonial (1965-), and precolonial (c. 13\textsuperscript{th}-15\textsuperscript{th} centuries), the former representing activity related to the ancient settlement of Temasek-Singapura. The artefacts found within this early layer (95-175 cmbd in each trench) primarily comprised two types of ceramic sherds: imported Chinese porcelain and stoneware dating between the 11\textsuperscript{th}-14\textsuperscript{th} centuries \cite{Lim_2012} and locally-made and imported earthenware vessels; the former form the focus of this study (see below). Chinese coins minted from the Song (960-1279 CE) and Yuan dynasties (1279-1368 CE), glass beads and bangles, worked stone, shellfish and bone remains, and an impressive record of charcoal are included in this precolonial artefact assemblage \cite{Miksic_and_Lim_2003}.

\subsection{Methodology: unsupervised clustering of image features}

Ceramic sherd data is provided via images of the fragments of ceramic sets stored at the Institute of Southeast Asian Studies (ISEAS) of the National University of Singapore, where they were grouped by visual inspection by the third author. In this work, we analyse five different ceramic sets composed of a varying number of pieces, as reported in Table \ref{tab:Ceramic_Sets}. The images are pre-processed following the same steps used in the numismatic application (see Section \ref{sec:Coins_preprocessing}), with the only difference that the images are now resized to $1000 \times 1000$ pixels. The aim of this analysis is to extract features form the available images and exploit them to recover the known grouping into sets. While this is a simple task to perform with the ceramic sets at hand, it might pose as a challenging endeavour when a large number of images are to be analysed simultaneously.

\begin{table}[h!]
\begin{center}
     \begin{tabular}{cc}
      Set \# & Sherds \\\hline 
     \multirow{3}{*}{Set 1} & 
     \includegraphics[width=0.1\textwidth]{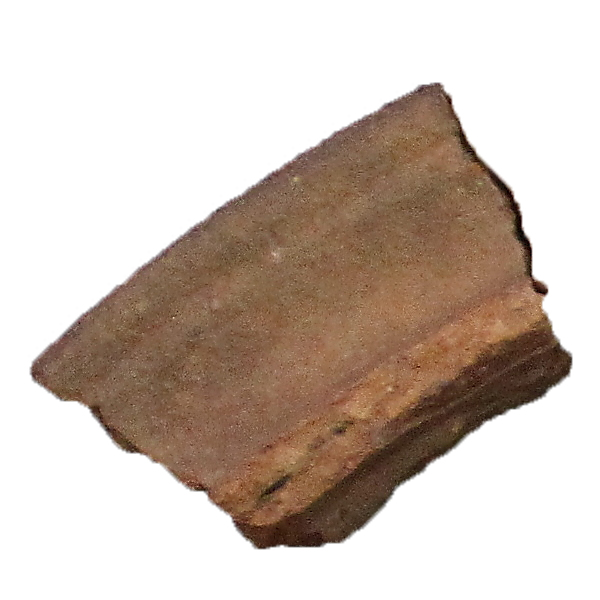}
     \includegraphics[width=0.1\textwidth]{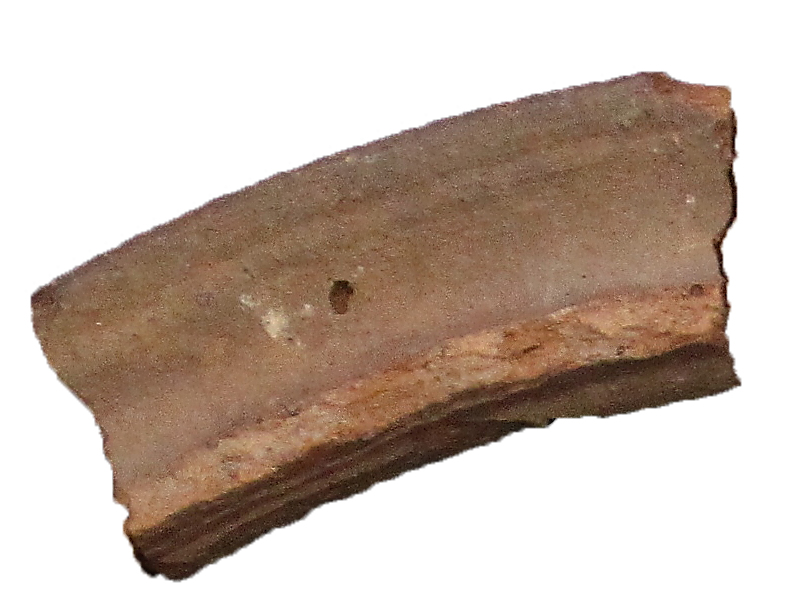}
     \includegraphics[width=0.1\textwidth]{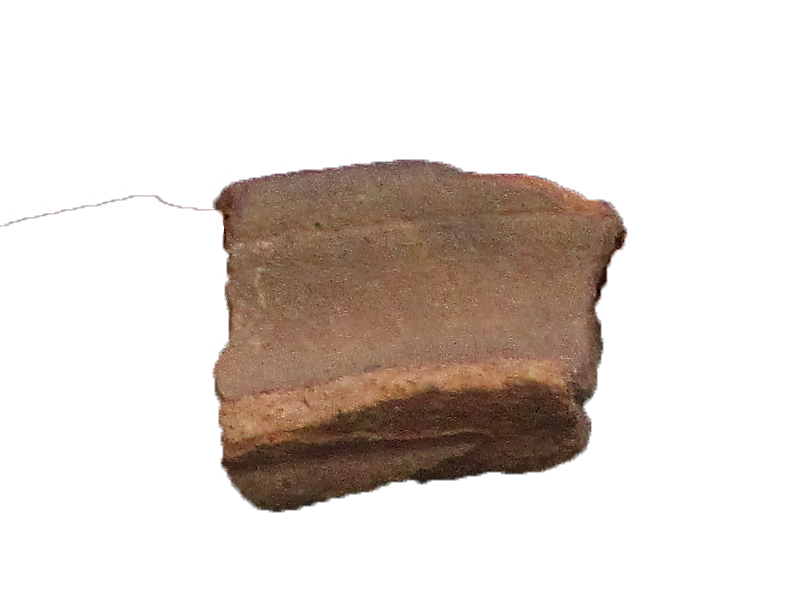}
     \\
     &
     \\
     & 
     \includegraphics[width=0.1\textwidth]{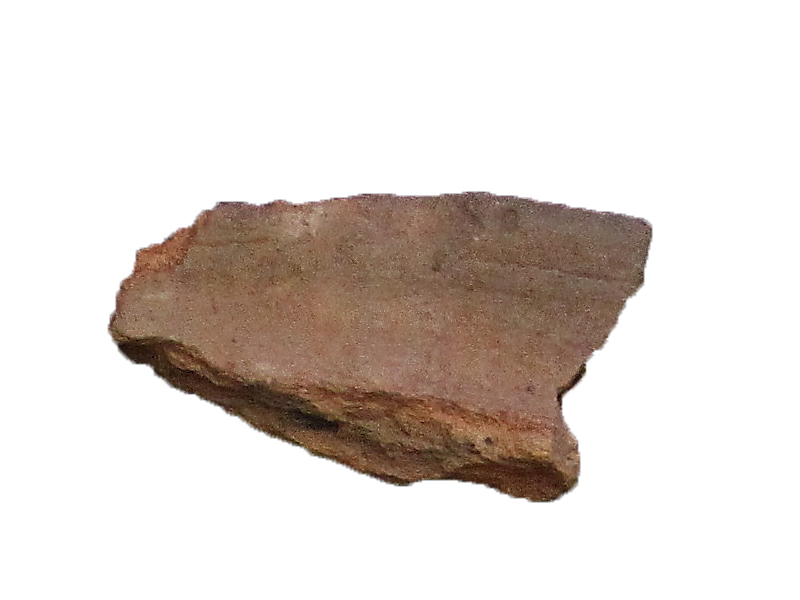}
     \includegraphics[width=0.1\textwidth]{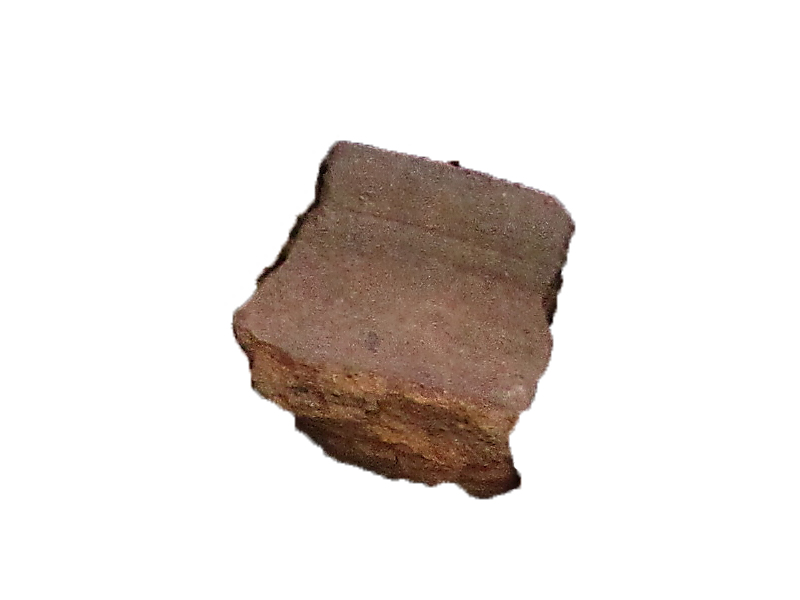}
     \includegraphics[width=0.1\textwidth]{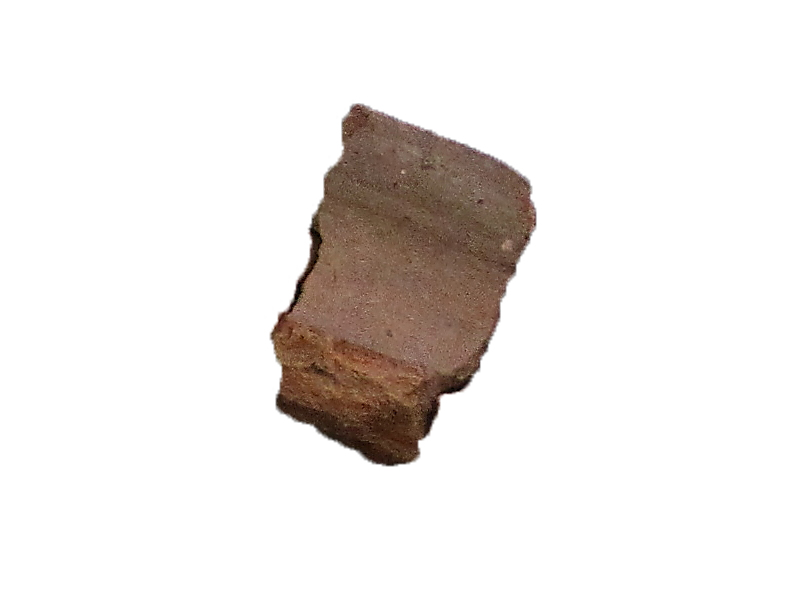}
     \includegraphics[width=0.1\textwidth]{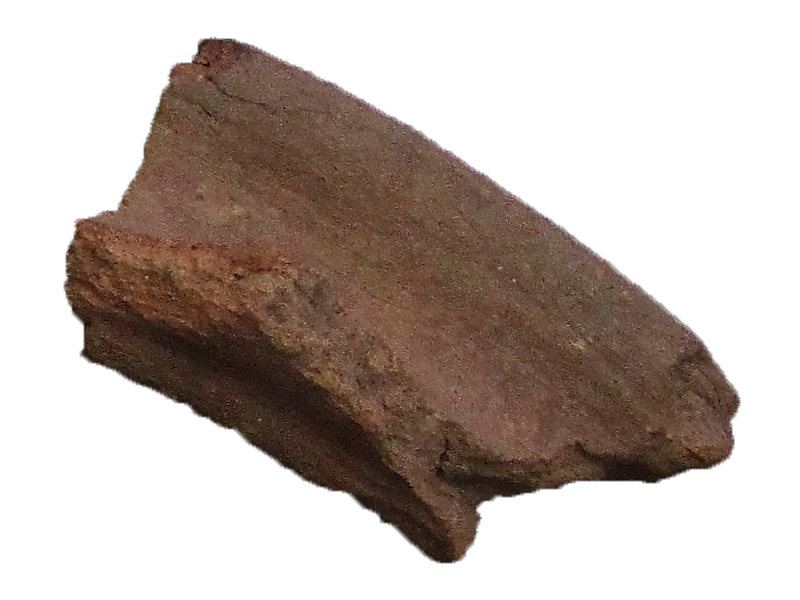}
     \\\hline
     \multirow{1}{*}[1.5ex]{Set 2} & 
     \includegraphics[width=0.1\textwidth]{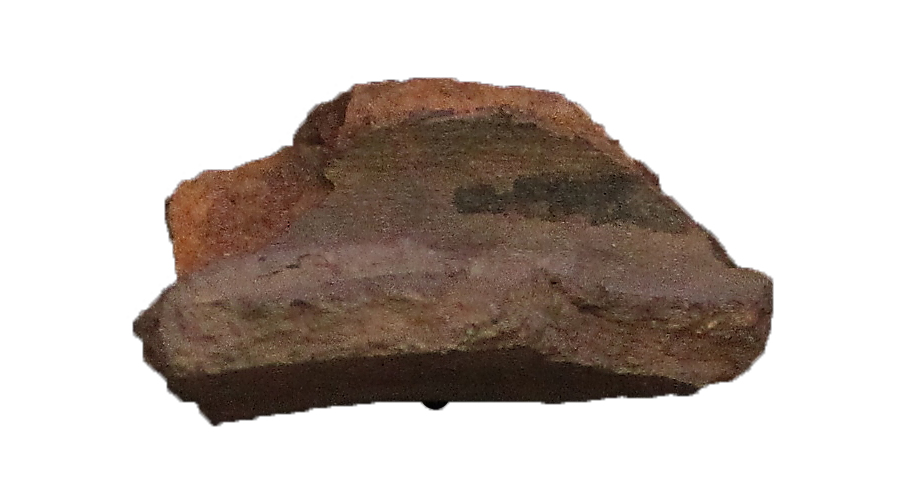}
     \includegraphics[width=0.1\textwidth]{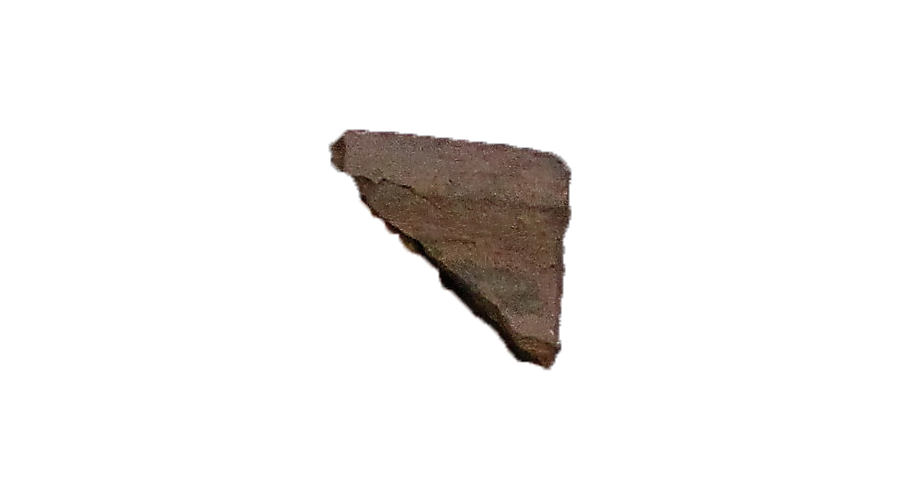}
     \\\hline
     \multirow{1}{*}[1.5ex]{Set 3} & 
     \includegraphics[width=0.1\textwidth]{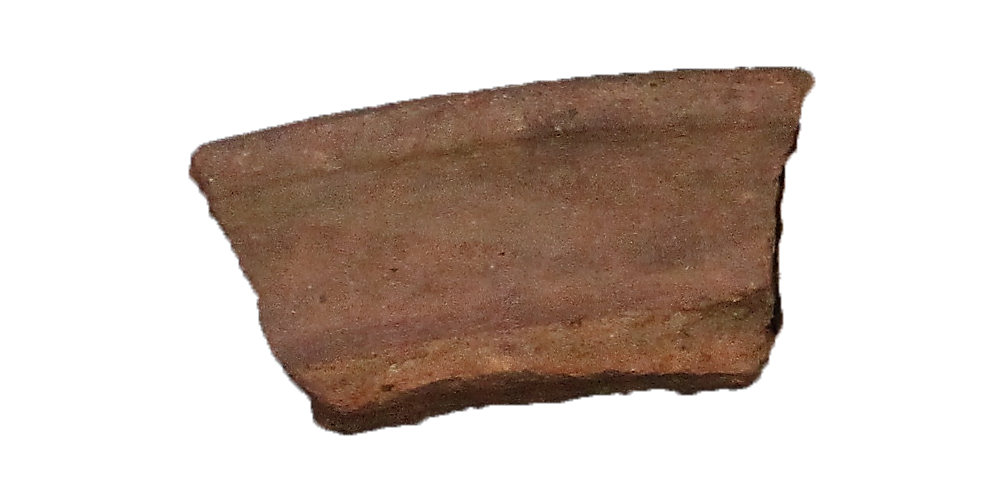}
     \includegraphics[width=0.1\textwidth]{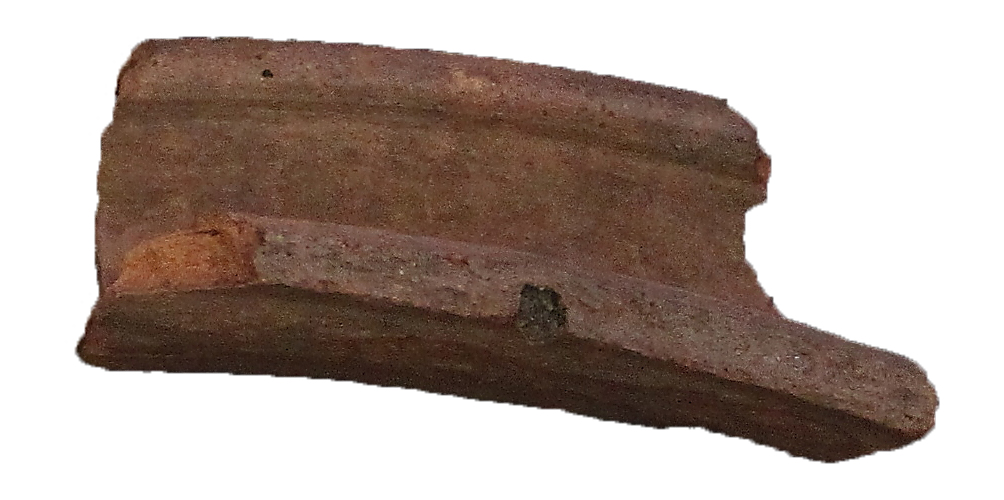}
     \\\hline
     \multirow{1}{*}[1.5ex]{Set 4} & 
     \includegraphics[width=0.1\textwidth]{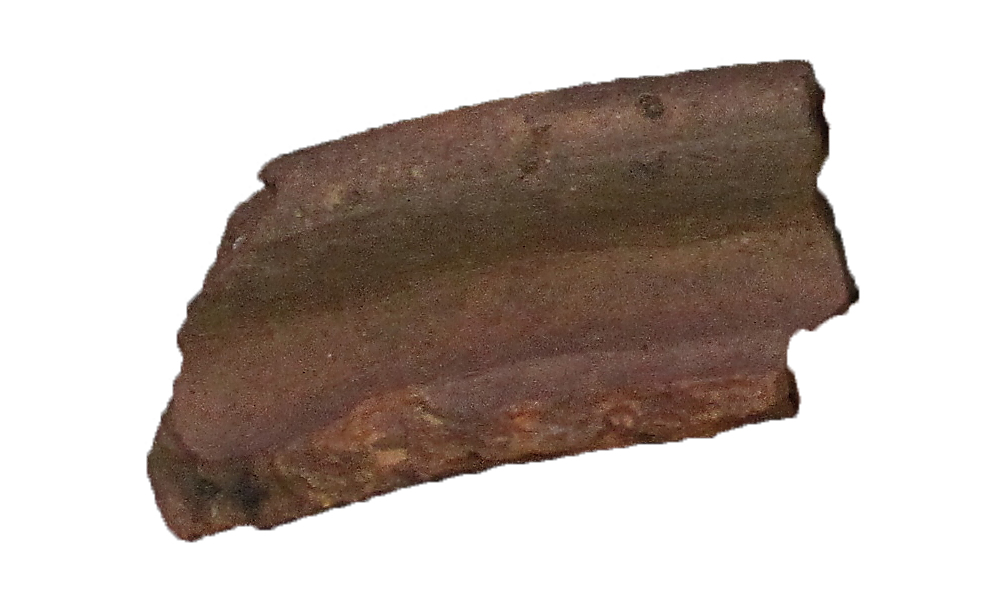}
     \includegraphics[width=0.1\textwidth]{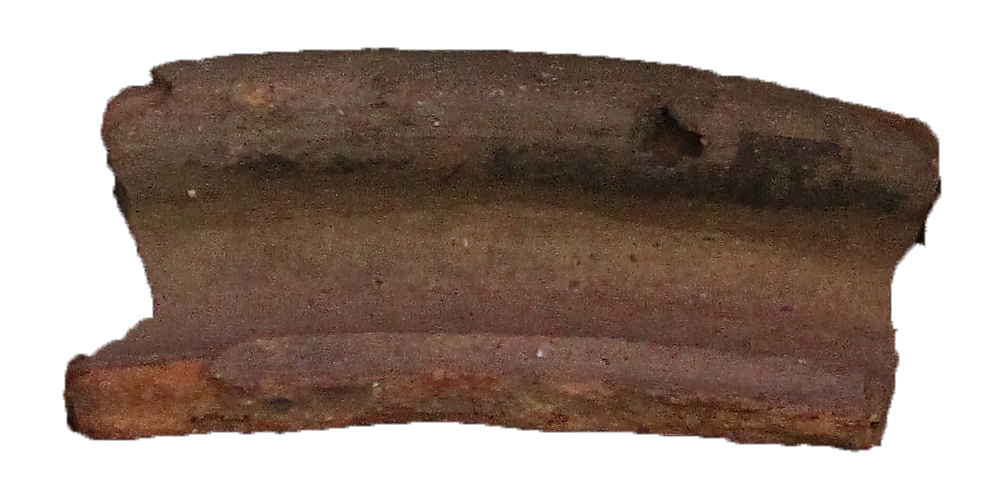}
     \includegraphics[width=0.1\textwidth]{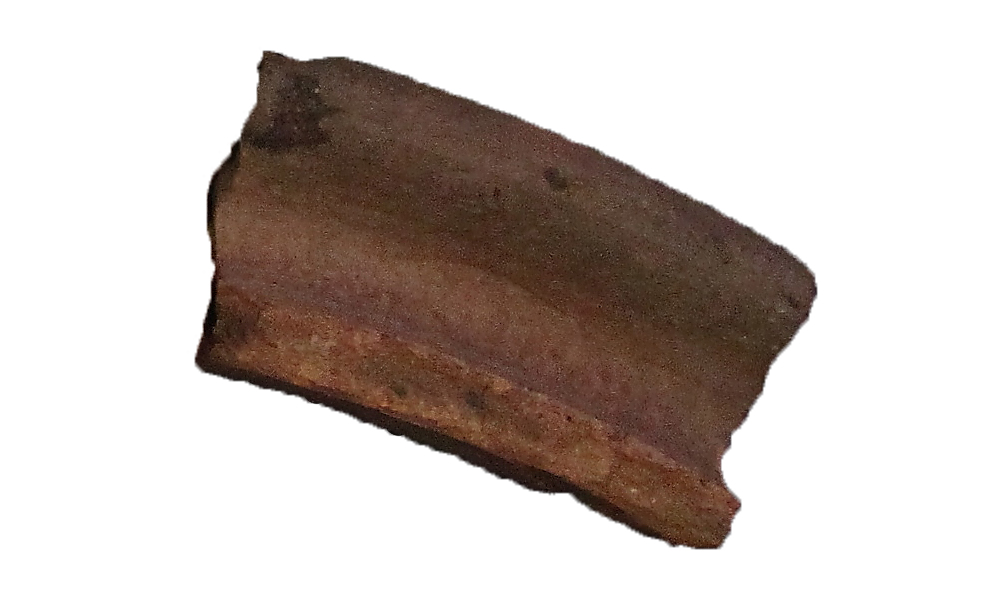}
     \\\hline
     \multirow{1}{*}[1.5ex]{Set 5 back} & 
     \includegraphics[width=0.1\textwidth]{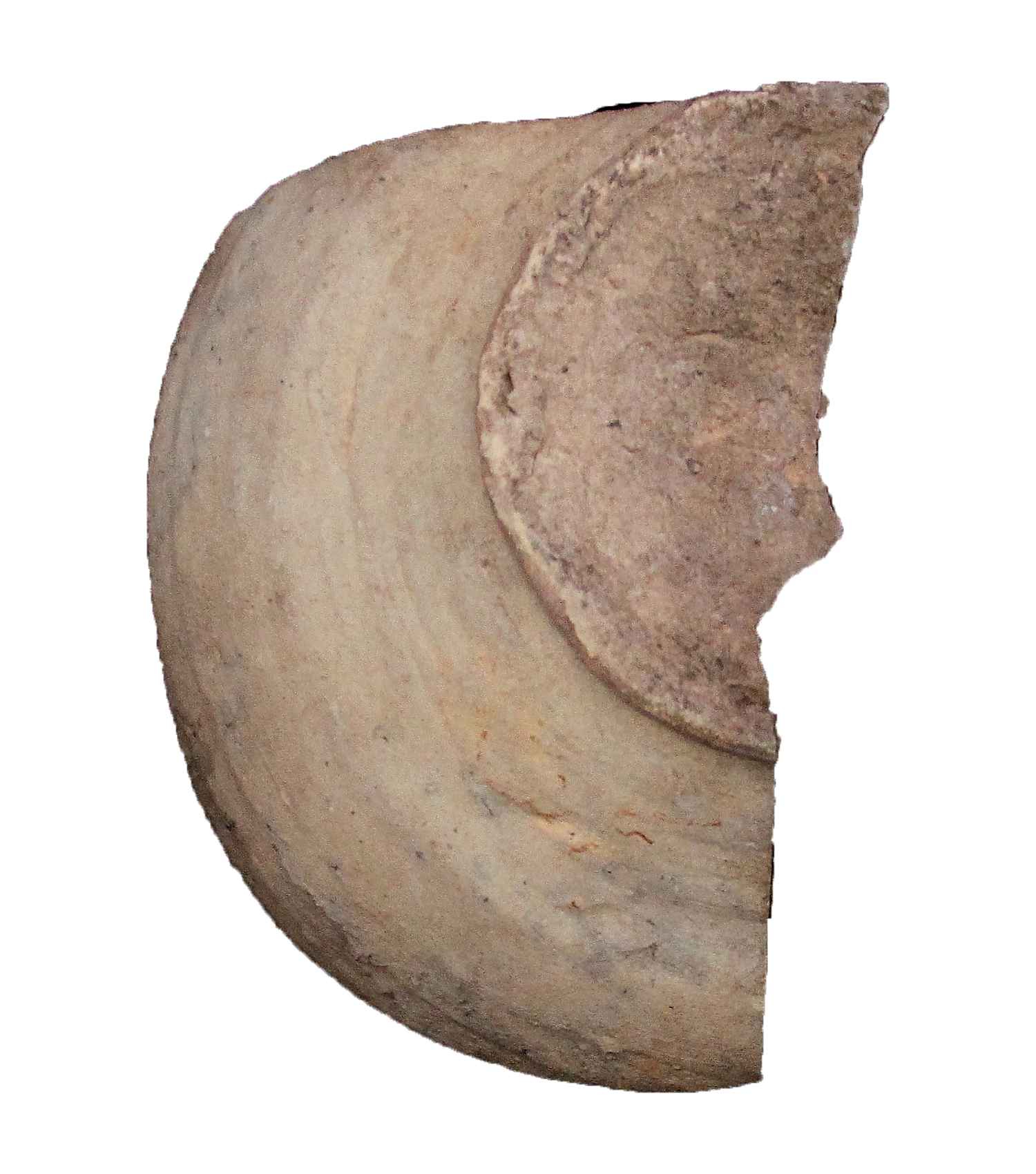}
     \includegraphics[width=0.1\textwidth]{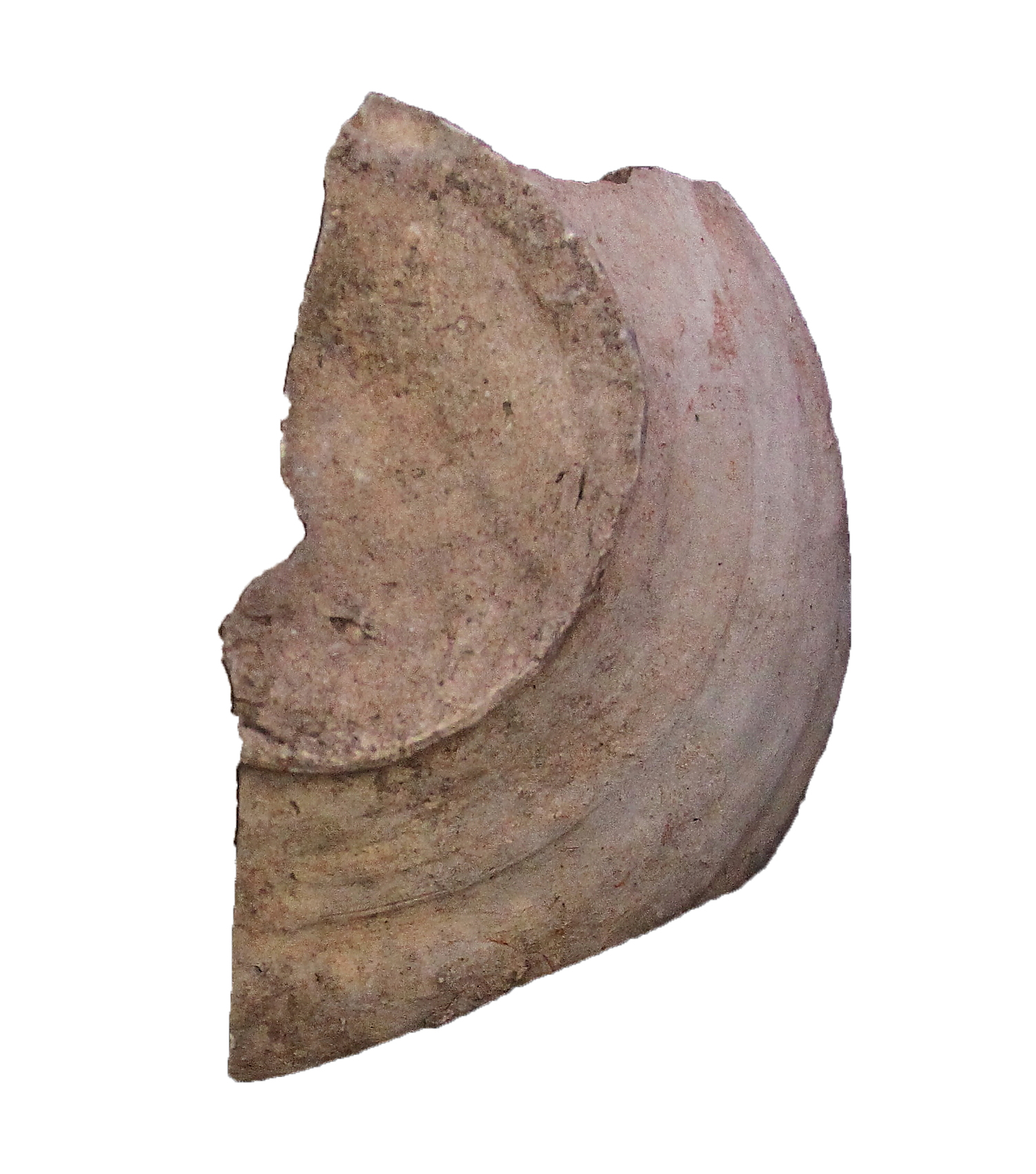}
     \\\hline
     \multirow{1}{*}[1.5ex]{Set 5 front} & 
     \includegraphics[width=0.1\textwidth]{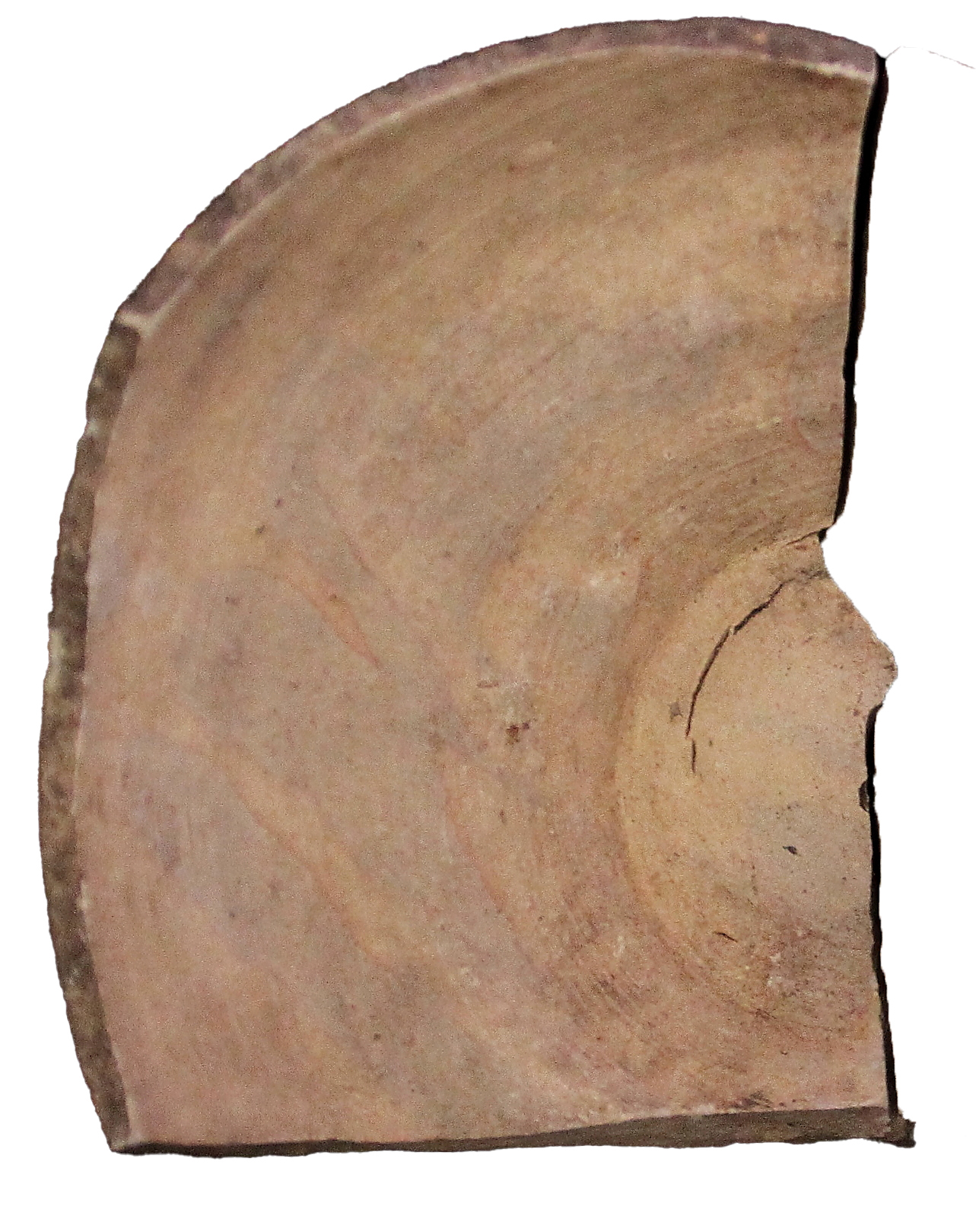}
     \includegraphics[width=0.1\textwidth]{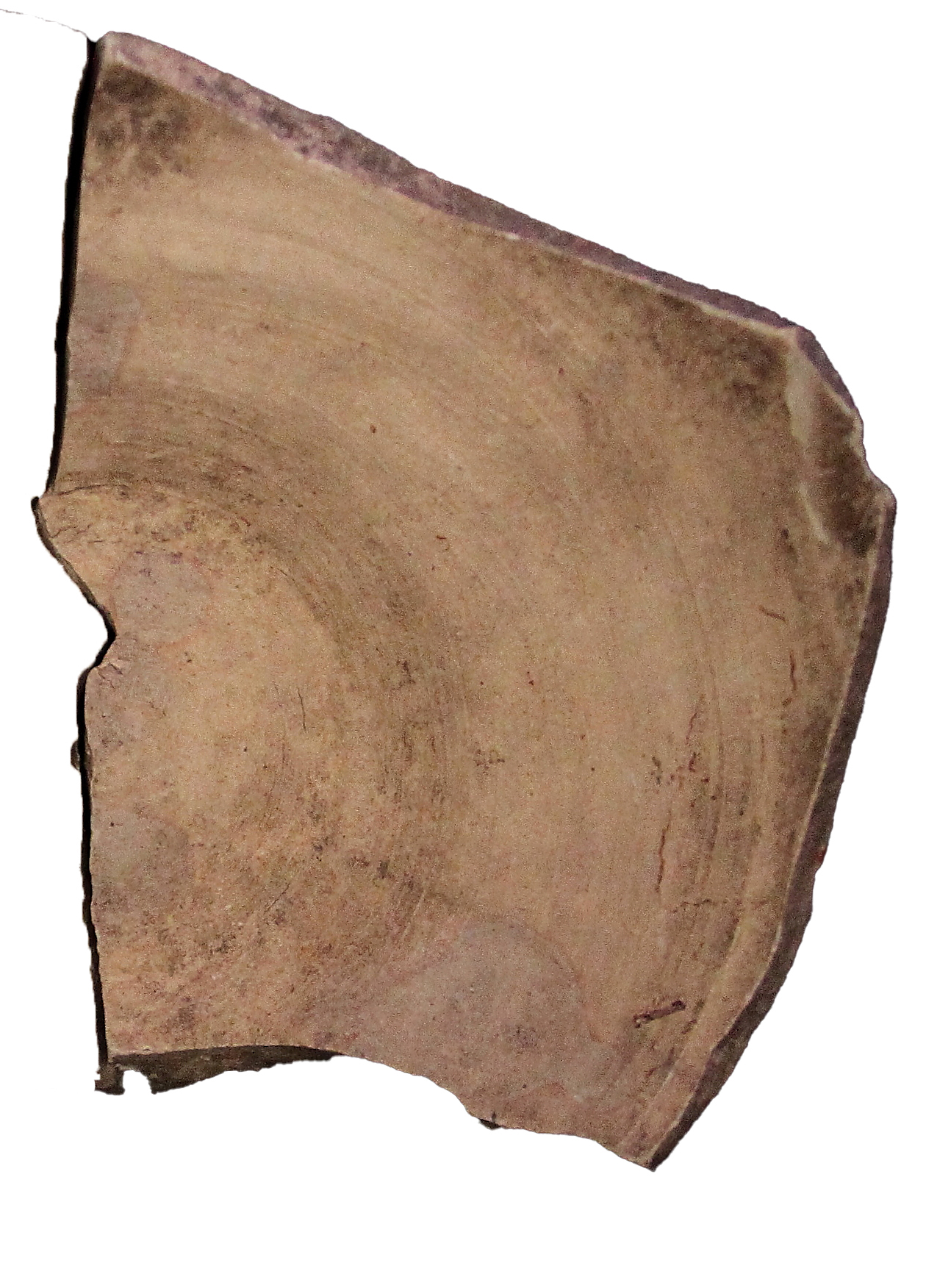}
     \\\hline
     \end{tabular}
     \caption{Images of the ceramics sherds within each set.}
     \label{tab:Ceramic_Sets}
\end{center}
\end{table}

\begin{figure}[ht]
	\centering
     \includegraphics[width=0.45\textwidth]{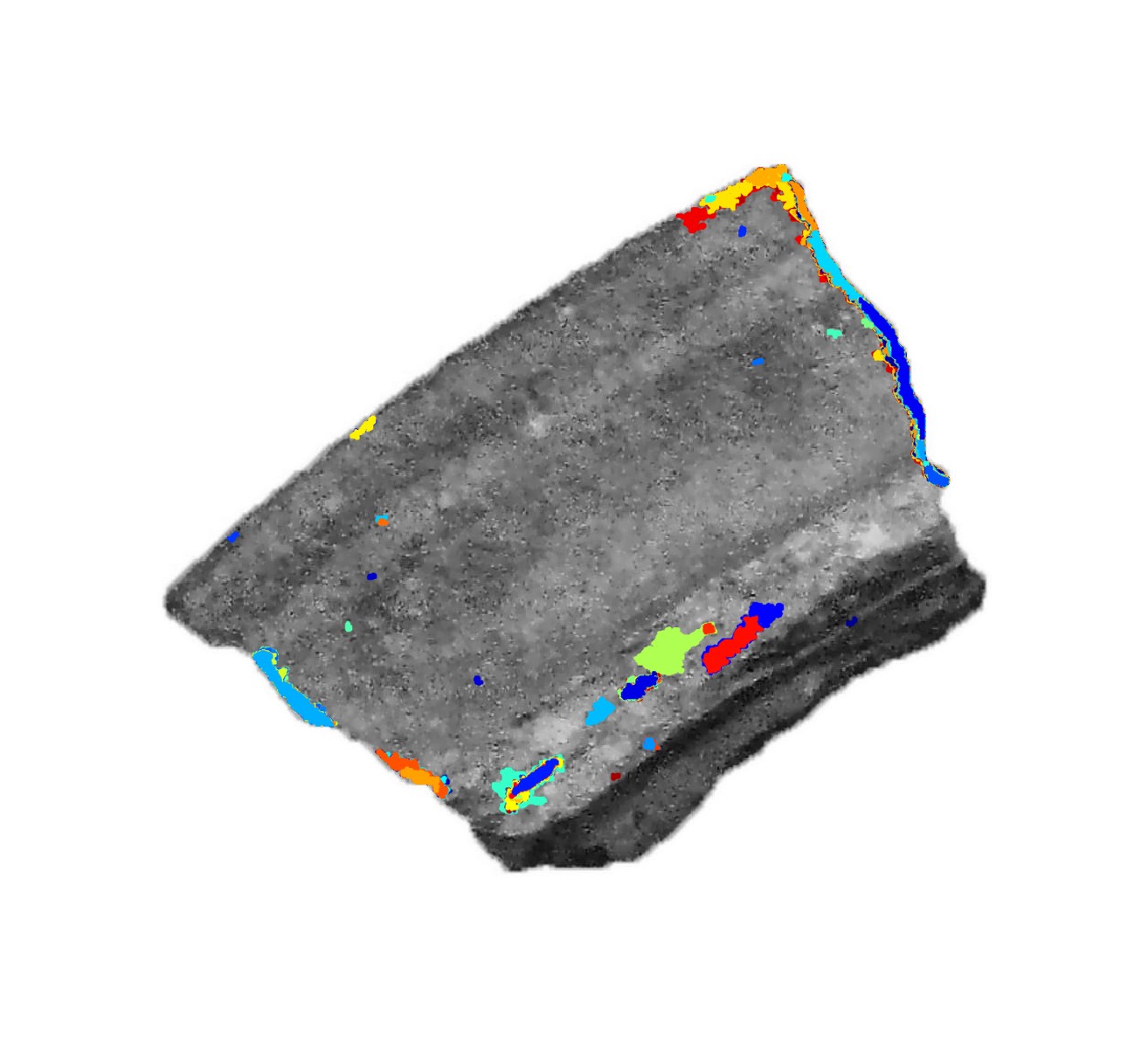}
    \includegraphics[width=0.45\textwidth]{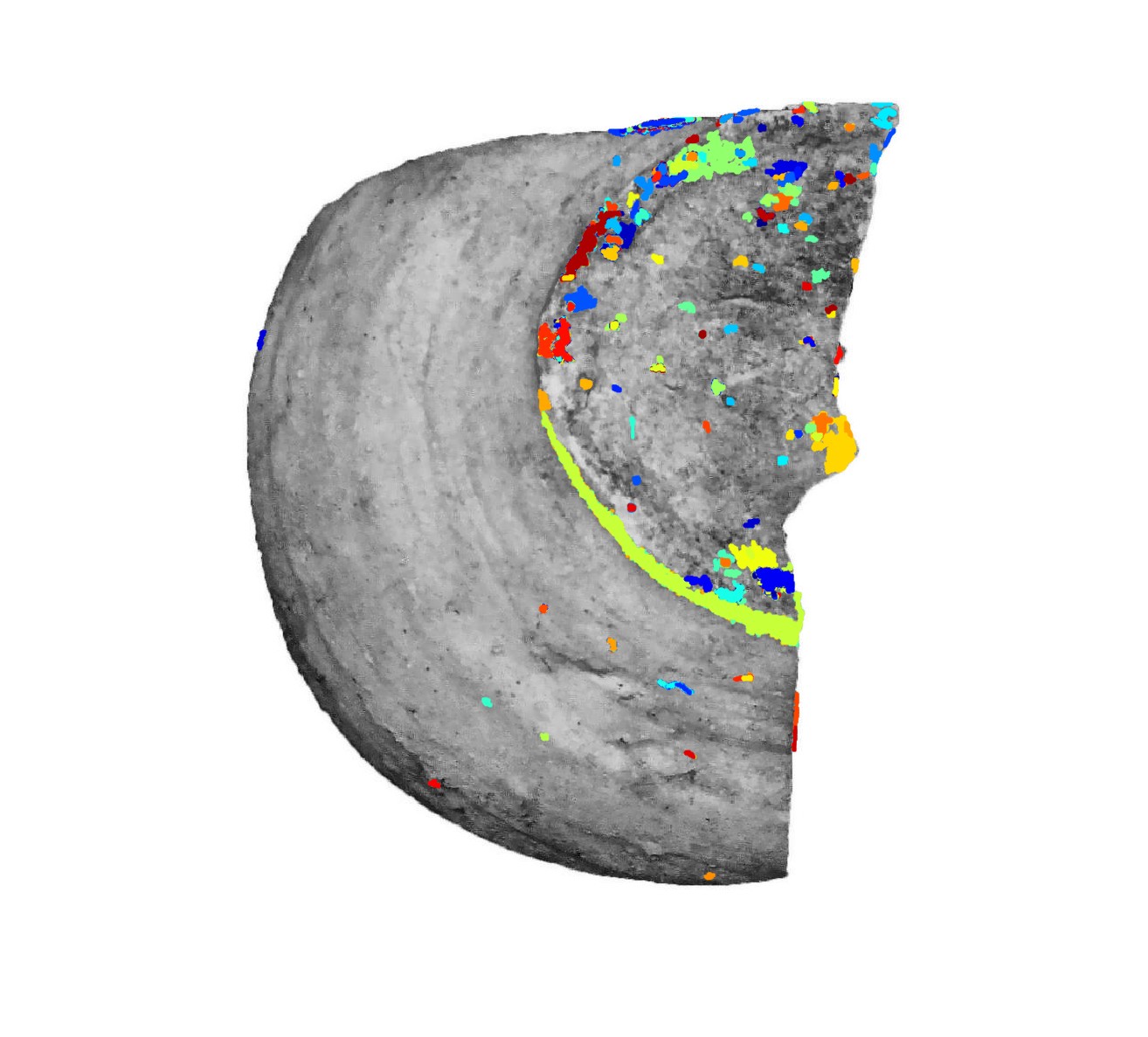}
    \\
	\includegraphics[width=0.45\textwidth]{IMG_0031_Sherd_1_Part_1.jpg}
    \includegraphics[width=0.45\textwidth]{IMG_0226_Part_1.jpg}
	\caption{Top row: MSER connected components identified as circles. Bottom row: top 10\% strongest features extracted by MSER for further analysis.}
	\label{fig:MSER_regions}
\end{figure}

After pre-processing, the enhanced grey-scale images are analysed using feature detection and extraction methods, available within \texttt{Matlab}'s \textit{Computer Vision Toolbox}. We use the Maximally Stable Extremal Regions (MSER) algorithm \cite{Matas_etal_2002} to detect relevant features from each image. MSER extracts as features the connected components of the sets of pixels in the image characterised by similar intensity. The algorithm only selects the stable regions among the connected components, by increasing the region size and checking that the incremental differences are smaller than a threshold provided by the user. The centres of these regions are then used for feature extraction. This process can be visualised as follows. Firstly,  an additional third dimension is added to the image representing each pixel's intensity, yielding a 3D image object. Thus,  we perform a ``flooding'' of the image with water by steady increments. At each step, the regions connected by water represent the connected components of the image that, if stable enough, can be extracted as features. Examples of connected regions and extracted features identified by the MSER algorithm are shown in Figure \ref{fig:MSER_regions} for some of the ceramic sherds available. In this work, we consider only the top 10\% strongest features of each image for further analysis.

\begin{figure}[ht]
	\centering
     \includegraphics[width=1\textwidth]{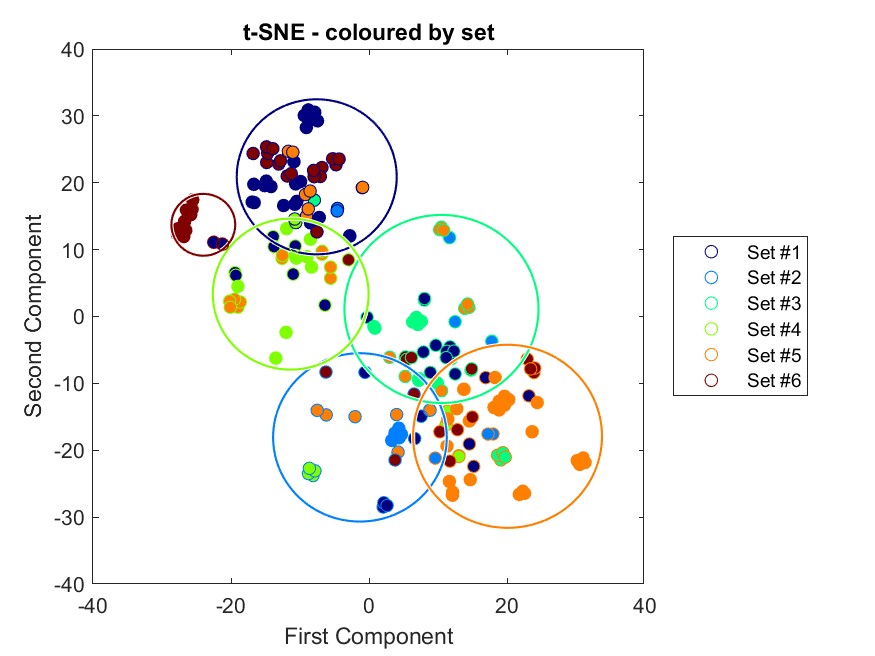}
	\caption{t-SNE projection of the features extracted from the ceramic images into a two-dimensional space. For each point representing a feature, the inner colour refers to the available information regarding the corresponding image (i.e., which ceramic set or which ceramic sherd the feature belongs to), while the outer shape colour refers to the result of the $k$-means clustering algorithm. The circles in each image contain all the features belonging to the same $k$-means cluster.}
	\label{fig:tSNE_kmeans}
\end{figure}

As a result of the feature extraction procedure, we obtain a dataset of $213 \times 64$ features to use for clustering purposes. The dimension of the feature matrix results from standard feature extraction techniques. The next step in the analysis is to identify groups of images using the information contained in the feature matrix. To do so, we resort to the popular $k$-means algorithm \cite{MacQueen_1967} and compare the result with the information arising from visual inspection. Due to the high dimension of the problem, we perform an initial dimension reduction step to improve the performance of the clustering algorithm. Specifically, we opt for the t-Stochastic Neighbour Embedding (t-SNE) approach \cite{Van_der_Maaten_2008}. t-SNE is a dimension reduction technique that estimates the data's optimal projection in two or three dimensions. This techniques is widely used in high-dimensional problems, such as single-cell studies, to visualise patterns within the data in a lower dimensional setting. We apply the t-SNE approach directly to the feature matrix to obtain a two-dimensional representation of the feature vectors. Finally, we apply the $k$-means algorithm to cluster the projected features. The number of clusters in the $k$-means algorithm is pre-specified by the user, and here we fix it to the number of observed ceramic sets, i.e. 5. We show in Figure \ref{fig:tSNE_kmeans} the results of the cluster analysis and compare them with the information available regarding the ceramic images. We plot the features extracted from the ceramic images and projected into a two-dimensional space via t-SNE. Each point represents a row of the feature matrix after t-SNE projection. Some of the images are characterised by features that are in close proximity in the t-SNE space and are therefore more easily clustered together. On the other hand, some images' features are located far apart and are therefore assigned to different clusters. Nonetheless, the method achieves an accuracy, as measured by the Rand index, of the 0.7029 when compared to the set information. 

\subsection{Methodology: vessel reconstruction}
\label{sec:2Drec}

In this section, we describe an automatic method to match sherds with the ultimate goal of reconstructing a ceramic object. We exploit computer vision methods which allow to retrace the contour of each sherd and represent it as a  2D image. Ceramic reconstruction though  computer images  has received increasing interest in recent years, with the development of several computational strategies in support of historical research, see \cite{Eslami_etal_2020} for an extensive review on the topic. More recently, \cite{Rasheed_and_Nordin_2015,Eslami_etal_2021} proposed a curve fitting approach to detect matching borders in images of broken ceramics. The 2D images are obtained through high-definition photographs taken along the vertical axis of  each sherd, i.e. on the side where the pottery is broken. The shape of these edges are then extracted by exploiting the Canny
filter algorithm \cite{Ding_and_Goshtasby_2001}, a method used for border detection in binary 2D images. The outline of the shapes, typically represented as a closed ellipsoid line, is then split into two parts, each representing a different curve to be analysed. Either a polynomial curve \cite{Rasheed_and_Nordin_2015} or wavelets  \cite{Eslami_etal_2021} are fitted to these curves with the goal of both noise removal and main feature extraction. Finally, matching of the curves, and therefore of the edges of the sherds, is  performed by comparing the resulting curve fittings.
 
\begin{figure}[ht!]
	\centering
	\includegraphics[width=1\textwidth]{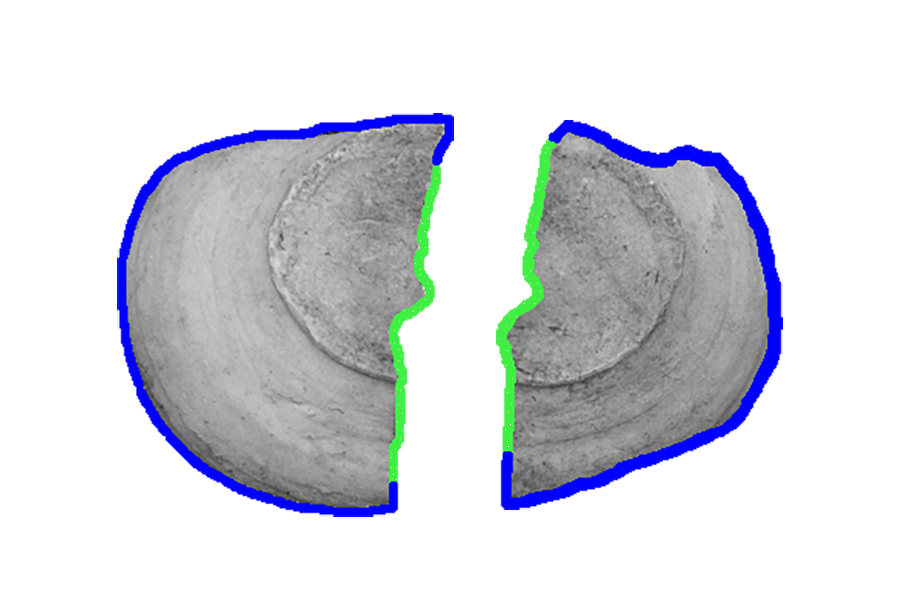}\\
    \includegraphics[width=1\textwidth]{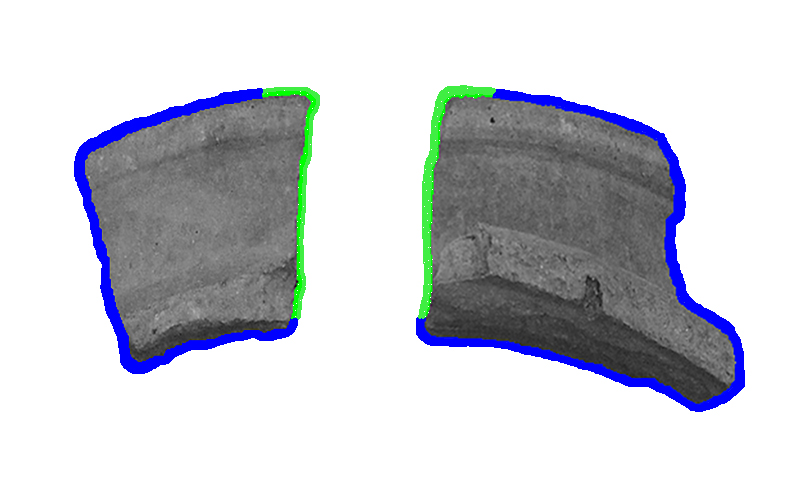}
	\caption{Matched sherd pairs obtained through automatic vessel reconstruction. Green lines represent matched edges.}
	\label{fig:Matched_sherds}
\end{figure}

In this work, we opted for a different approach, inspired by solving a jigsaw puzzle. This is motivated by the fact that available 2D images were taken perpendicularly with respect to the breaking point of the sherds, thus making it impossible to apply the methods described above. We proceeded as follows. We first applied the Canny filter for the detection of the contour of each sherd image. Then, for each pair of images within a set, we compared segments of the corresponding contours (sub-contours) by using a moving window of pixels. Each pair of sub-contours was compared following an approach analogous to that found at the GitHub repository \texttt{https://github.com/MaximTerleev/Jigsaw-Puzzle-AI}. In particular, we computed the seven invariants based on Hu moments \cite{Hu_1962} of each sub-contour. These quantities describe important image features and are invariant to translation, scale, and rotation. We  used them to produce different dissimilarity measures between each pair of sub-contours, as suggested by \texttt{https://learnopencv.com/shape-matching-using-hu-moments-c-python/}. Differently from previous approaches, we also included the seventh invariant, which provides information about mirror images. By minimising the sum of the squared dissimilarities for each pair of sub-contours, we were able to identify the most promising matches for where two sherds might align. More in details, for each pair of sherds, we inspected the top three sub-contour matches, and were able to correctly identify pairs of sherds to be matched together as well as the exact location on the edge. See Figure \ref{fig:Matched_sherds} for some examples of automatic vessel reconstruction.

\clearpage

\subsection{Discussion}

Automatic methods for analysing and reconstructing ceramic sherds allow for a more comprehensive and wider-scale interpretation of the types of ceramic objects constructed and uses within early societies, especially in the absence of ancient literary sources describing these cultures in great detail. The establishment of a digital open-access archive of ceramic artefacts, too, can facilitate future both intra- and inter-site comparisons of ceramic artefacts, which would supplement existing cross-cultural and contemporary ethnographic studies. For instance, various studies chronicling ceramic production across modern mainland Southeast Asia highlight the use of either a paddle and anvil or a spatula to smoothen and shape a vessel into its final pre-firing form; a ‘slow’ wheel, intermittently spun by hand to shape the vessel body, may be used in the shaping process as well, but not in the same manner as the faster potter’s wheel \cite{Lefferts_and_Cort_2003}. The absence of rilling (decoration comprising fine incised close-set lines) on local coarse and medium-tempered earthenware ceramics excavated from the STA site suggests a similar practice occurred in 14th century Temasek \cite{Lim_2012}, which has been suggested as a product of precolonial Singaporean potters of a Malay-Nusantao or Malay-speaking people \cite{Lim_2012}. A database of reconstructed or semi-reconstructed ceramics, in this case earthenware vessels from the STA site, would help supplement or challenge these theories while also providing valuable information on the daily lives of early Singapore's residents, the island's ethnic diversity, and the longevity of various methods of ceramic production across Southeast Asia. 

Moreover, computer vision tools can be used to recover and  typologise different decorative motifs, which could be, for example, used to reconstruct the art-historical evolution of specific civilisations, enabling also a faster and more comprehensive comparison across cultures.     

\section{Conclusions}
This work highlights the increasing potential of  statistical methods in Digital Humanities when applied to archaeological datasets, underscoring archaeology's position at the forefront of technological innovation within the realms of humanities and social sciences.  Rather than portraying digital archaeology as deceptively straightforward, normative, de-skilling, or automating, it should be seen as an opportunity to foster meaningful engagement. These shicstudies, it is important to note, supplement existing knowledge bases, and do not serve to substitute the hard work undertaken to contextualise the datasets of artefacts involved\cite{Morgan_2022}. 

Digital Humanities represent a remarkable fusion of tradition and technology, revolutionising, for instance,  the way we approach the study and appreciation of coins and ceramics. As this field continues to evolve, it holds promise to further our understanding of history, culture, and economics through the lens of ancient artefacts. However, it is  important to note that, while digital humanities holds great potential for discovery and research, there exist challenges to be overcome, such as standardisation of data formats, the need for robust metadata protocols and issues related to the authenticity and integrity of digital records \cite{Gruber_Meadows_2021}. Emerging automated learning-based techniques, meanwhile, provide a valuable tool for interpreting the wealth of archaeological image-based data (photographs, 3D scans, or even hand-drawings) accumulated over more than a century of interpretation. 

Digital methods of analysis allow for the enhancement of various tasks in archaeological analysis while opening the door to broader research possibilities. However, making data accessible and suitable for these methods often necessitates labour-intensive and costly processes of data structuring and annotation, which in turn are still subordinated by more traditional methods of artefact analysis, for example manual drawing and sorting. Therefore, a key requirement for the future of digital artefact analysis is the development of  more efficient methods for computer-assisted preparation and annotation of large datasets of artefacts. The recent studies highlighted in this survey have set the stage for pursuing such objectives in the future\cite{Karl_etal_2022}.

\pagebreak

\section*{Acknowledgements}

We would like to thank Dr. Blaise Kilian and associated staff of the SOSORO Museum of Economy and Money in Phnom Penh, Cambodia, as well as Dr. Hoang Anh Tuan and associated staff from the Ho Chi Minh City History Museum, for permissions to examine, measure, and photograph all "Rising Sun" coins used in this study. Special thanks to Tran Ky Phuong for helping attain permissions in Vietnam, and for Robert Wicks for his initial assessment of our methodology and field collection methods. This paper was written under Grant No. MOE-T2EP40121-0021 from the National University of Singapore.



\begin{thebibliography}{9}

\bibitem{Aagard_and_Marcher_2015} Aagaard S and M Marcher (2015) \textit{The Microscope Drawing Tube Method (MTDM) – An Easy and Effective Way to Make Large-Scale Die Studies}. The Numismatic Chronicle, 175, 249-262.

\bibitem{Van_Alfen_2017} Van Alfen P (2017) \textit{The Computer Aided Die Studies Program}. URL: http://numismatics.org/pocketchange/cads/

\bibitem{Anwar_etal_2015} Anwar H, S Zambanini and M Kampel (2015) \textit{Coarse-grained ancient coin classification using image-based reverse side motif recognition}. Machine Vision and Applications, 26(2-3), 295-304.

\bibitem {Aslan_etal_2020} Aslan S, S Vascon and M Pelillo (2020) \textit{Two sides of the same coin: Improved ancient coin classification using Graph Transduction Games}. Pattern Recognition Letters, 131, 158-165.

\bibitem{Aycock_2021} Aycock J (2021) \textit{The Coming Tsunami of Digital Artefacts}. Antiquity, 95, 1584-9.

\bibitem{Beaujard_2019}Beaujard P (2019) \textit{Chapter 11: Southeast Asia, An Interface Between Two Oceans}. The Worlds of the Indian Ocean: A Global History, Volume 1 – From the Fourth Millennium BCE to the Sixth Century CE. Cambridge University Press, 518.

\bibitem{Bentkowska_MacDonald_2018} Bentkowska-Kafel A and L MacDonald (2018) \textit{Digital techniques for documenting and preserving cultural heritage} (p. 370). Arc Humanities Press.

\bibitem{Berry_2011} Berry DM (2011) \textit{The COmputational Turn: Thinking about the Digital Humanities}. Culture Machine, 12, 1-22.

\bibitem{Berry_2019} Berry DM (13th Feb 2019) \textit{What are Digital Humanities?} The British Academy Blog. Retrieved from: https://www.thebritishacademy.ac.uk/blog/what-are-digital-humanities/

\bibitem{Binford_1965} Binford LR (1965) \textit{Archaeological Systems and the Study of Culture Process}. American Antiquity, 31(2), 203-210.

\bibitem{Betancourt_etal_2022} Betancourt B, Zanella G, Steorts RC (2022) \textit{Random Partition Models for Microclustering Tasks}. Journal of the American Statistical Association, 117(539), 1215-1227.

\bibitem{Breier_2010} Breier M (2010) \textit{GIS for Numismatics–Methods of Analyses in the Interpretation of Coin Finds}. In Mapping Different Geographies (pp. 171-182). Berlin, Heidelberg: Springer Berlin Heidelberg.

\bibitem{Brown_etal_2008} Brown, B.J, C Toler-Franklin, D Nehab, M Burns, D Dobkin, A Vlachopoulos, C Doumas, S Rusinkiewicz, and T Weyrich, (2008)
\textit{A system for high-volume acquisition and matching of fresco fragments: Reassembling Theran wall paintings}. ACM Transactions on Graphs, 27, 1-9.

\bibitem{DeCallatay_1995} De Callatay E (1995) \textit{Calculating Ancient Coin Production: Seeking a Balance}. The Numismatic Chronicle, 155, 289-311.

\bibitem{Cappon_1886} Cappon P (1886) \textit{Trouvaille de monnaises en Cochinchine}. Revue Numismatique, 3, vols. 4-5, 295-297.

\bibitem{Carter_etal_2021} Carter AK, Dussubieux L, Stark MT, Gilg HA (2021) \textit{Angkor Borei and Protohistoric Trade Networks: A View from the Glass and Stone Bead Assemblage}. Asian Perspectives, 60(1), 32-70.

\bibitem{Chang_2016} Chang KT (2016) \textit{Geographic information system}. International encyclopedia of geography: people, the earth, environment and technology, 1-10.

\bibitem{Clarke_1973} Clarke D (1973) \textit{Archaeology: The Loss of Innocence}. Antiquity, 47(185), 6-18.

\bibitem{Coedes_1968} C\oe d\`{e}s (G 1968) \textit{The Indianized States of Southeast Asia}. Smithies, M trans. Honolulu, HI: University of Hawaii Press.

\bibitem{Cooper_and_Arandjelovic_2019} Cooper J and Arandjelovi\'{c} O (2019) \textit{Understanding ancient coin images}. In INNS Big Data and Deep Learning conference. Springer, Cham, 330-340.

\bibitem{Cooper_and_Arandjelovic_2020} Cooper J and Arandjelovi\'{c} O (2020) \textit{Learning to Describe: A New Approach to Computer Vision Based Ancient Coin Analysis}. Sci, 2(2), 27.

\bibitem{Cribb_and_Bracey_2019} Cribb J and Bracey R (2019) \textit{Kushan Coins: A Catalogue Based on the Kushan, Kushano-Sasanian and Kidarite Hun Coins in The British Museum, 1st-5th Centuries AD}. London: British Museum Press.

\bibitem{Coedes_1968} C\oe d\`{e}s (G 1968) \textit{The Indianized States of Southeast Asia}. Honolulu, HI: University of Hawaii Press.

\bibitem{Ding_and_Goshtasby_2001} Ding L and Goshtasby A (2001) \textit{On the Canny edge detector}. Pattern recognition, 34(3), pp.721-725.

\bibitem{Dusmanu_2019a} Dusmanu M et al. (2019a) \textit{D2-Net: A Trainable CNN for Joint Detection
and Description of Local Features}. arXiv: 1905.03561 [cs.CV].

\bibitem{Dusmanu_2019b} Dusmanu M et al. (2019b) \textit{D2-Net: A Trainable CNN for Joint Detection and Description of Local Features}. In: Proceedings of the 2019 IEEE/CVF Conference on Computer Vision and Pattern Recognition.

\bibitem{Drucker_2021} Drucker J (2021) \textit{The Digital Humanities Coursebook: An Introduction to Digital Methods for Research and Scholarship}. New York, NY: Routledge.

\bibitem{Epinal_2013} Epinal G (2013) \textit{Quelques remarques relatives aux decouvertes monetaires d’Angkor Borei}. Numismatique Asiatique, Revue de la Societé de Numismatique Asiatique, 8, 31-43.

\bibitem{Epinal_and_Gardère_2014} Epinal G and Gard\`{e}re JD (2014) \textit{Cambodia from Funan to Chenla: A Thousand Years of Monetary History}. SOSORO Museum of Money and Economy: Phnom Penh, 93-125.

\bibitem{Eslami_etal_2020} Eslami D, Di Angelo L, Di Stefano P and Pane C (2020) \textit{Review of Computer-Based Methods for Archaeological Ceramic Sherds Reconstruction}. Virtual Archaeology Review, 11(23), 34-49.

\bibitem{Eslami_etal_2021} Eslami D, Di Angelo L, Di Stefano P, Guardiani E (2021) \textit{A Semi-Automatic Reconstruction of Archaeological Pottery Fragments from 2D Images Using Wavelet Transformation}. Heritage, 4(1), 76-90. 

\bibitem{Esty_2011} Esty W (2011) \textit{The Geometric Model for Estimating the Number of Dies}. Edipuglia, 43-58.

\bibitem{Esty_1986} Esty, W.W. (1986) \textit{Estimation of the size of a coinage: A survey and
comparison of methods.}The Numismatic Chronicle (1966-), 146, 185-215

\bibitem{Gao_etal_2019a} Gao T, Kovalsky SZ and Daubechies I (2019) \textit{Gaussian process landmarking on manifolds}. SIAM Journal on Mathematics of Data Science, vol. 1, no. 1, pp. 208–236, 2019.

\bibitem{Gao_etal_2019b} Gao T, Kovalsky SZ, Boyer DM and Daubechies I (2019) \textit{Gaussian process landmarking for three-dimensional geometric morphometrics}. SIAM Journal on Mathematics of Data Science, vol. 1, no. 1, pp. 237–267, 2019.

\bibitem{Goyal_1995} Goyal SR (1995) \textit{The Dynastic Coins of Ancient India}. Jodhpur, India: Kusumanjali Prakashan.

\bibitem{Graeber_2011} Graeber D (2011) \textit{Debt: The First 5000 Years}. Brooklyn, NY: Melville House, 320-321.

\bibitem{Grosman_2016} Grosman L (2016) \textit{Reaching the point of no return: the computational revolution in archaeology}. Annual Review in Anthropology, 45, 129-145.

\bibitem{Gruber_Meadows_2021} Gruber E and Meadows A (2021) \textit{Numismatics and Linked Open Data}. ISAW Papers.

\bibitem{Gutman_1976} Gutman P (1976) \textit{Ancient Arakan: With Special Reference to its Cultural History Between the 5th and 11th Centuries}. National Library of Australia.

\bibitem{Gutman_1978} Gutman P (1978) \textit{The Ancient Coinage of Southeast Asia}. Journal of the Siam Society, 66(1), 8-21

\bibitem{Hall_1999} Hall KR (1999) \textit{Coinage, Trade, and Economy in early South Asia and its Southeast Asian Neighbors}. The Indian Economic and Social History Review, 36(4), 431-459.

\bibitem{Heinecke_etal_2021} Heinecke A, Mayer E, Natarajan A, Jung Y (2021) \textit{Unsupervised Statistical Learning for Die Analysis in Ancient Numismatics}. arXiv, 2112.00290v.1 [es.CV] 1 Dec 2021.

\bibitem{Heng_2019} Heng D (2019) \textit{Premodern Island-Southeast Asian Hisotry in the Digital Age: Opportunities and Challenges through Chinese Textual Database Research}. Bijragen Tot De Taal-, Land- En Volkenkunde 175, 29-57.

\bibitem{Hess_etal_2018} Hess M, MacDonald LW and Valach J (2018) \textit{Application of multi-modal 2D and 3D imaging and analytical techniques to document and examine coins on the example of two Roman silver denarii}. Heritage Science, 6(1), 1-22.

\bibitem{Hu_1962} Hu MK (1962) \textit{Visual pattern recognition by moment invariants}. IRE transactions on information theory, 8(2), pp.179-187.

\bibitem{Kampel_and_Sablatnig_2003a} Kampel, M and R Sablatnig (2003a) \textit{An automated pottery archival and reconstruction system}. Journal of Visualization and Computer Animation, 14(3), 111-120.

\bibitem{Kampel_and_Sablatnig_2003b} Kampel, M and R Sablatnig (2003b) \textit{Rule based system for archaeological pottery classification}. Pattern Recognition Letters, 28(6), 740-747.

\bibitem{Kampel_and_Zaharieva_2008} Kampel M and Zaharieva M (2008) \textit{Recognizing ancient coins based on local features}. International Symposium on Visual Computing (pp. 11-22). Springer, Berlin, Heidelberg.

\bibitem{Karl_etal_2022} Karl S, Houska P, Lengauer S, Haring J, Trinkl E and Preiner R (2022) \textit{Advances in Digital Pottery Analysis}. it – Information Technology, 64(5), 195-216. https://doi.org/10.1515/itit-2022-0006

\bibitem{Klassen_etal_2018} Klassen S, Weed J and Evans D (2018) \textit{Semi-supervised machine learning approaches for predicting the chronology of archaeological sites: A case study of temples from medieval Angkor, Cambodia}. PLOS One, DOI: https://doi.org/10.1371/journal.pone.0205649

\bibitem{Khunti_2018} Khunti R (2018) \textit{he problem with printing Palmyra: exploring the ethics of using 3D printing technology to reconstruct heritage}. Studies of Digital Heritage, 2(1),1–12.

\bibitem{Lefferts_and_Cort_2003}Lefferts, L. and Cort, L.A. (2003) \textit{A Preliminary Cultural Geography of Contemporary Village-based Earthenware Production in Mainland Southeast Asia} Earthenware in Southeast Asia, Miksic, J., ed. Singapore: Singapore University Press, 300-10.

\bibitem{Liebert_1976} Liebert G (1976) \textit{Iconographie Dictionary of the Indian Religions}. Leiden.

\bibitem{Lim_2012} Lim TS (2012) \textit{14th Century Singapore: The Temasek Paradigm}. Thesis submitted for the title of Master of Arts, Department of Southeast Asian Studies, National University of Singapore.

\bibitem{Lipman_etal_2014} Lipman Y, Yagev S, Poranne R, Jacobs DW and Basri R (2014) \textit{Feature Matching with Bounded Distortion}. ACM Transactions on Graphics, 33(3). ISSN 0730-0301. doi: 10.1145/2602142.

\bibitem{vanLit_and_Morris_2024} van Lit LWC and Morris JH (2024) \textit{Digital Humanities and Religions in Asia: An Introduction}. Berlin, Boston: De Gruyter, 2024. https://doi.org/10.1515/9783110747607

\bibitem{Lowe_2004} Lowe D (2004) \textit{Distinctive Image Features from Scale-Invariant Keypoints}. International Journal of Computer Vision, 60, 91-110.

\bibitem{vanderMaaten_and_Postma_2006} van der Maaten L and Postma E (2006) \textit{Towards automatic coin classification}. Proc. of the EVAVienna, Vienna, Austria, 19–26.

\bibitem{Mahlo_2012} Mahlo D (2012) \textit{The Early Coins of Myanmar (Burma): Messengers from the Past – First Millennium AD}. Bangkok: White Lotus Press

\bibitem{Makarenkov_and_Lapointe_2004} Makarenkov V and Lapointe FJ (2004) \textit{A weighted least-squares approach for inferring phylogenies from incomplete distance matrices}. Bioinformatics, 20, 2113–2121.

\bibitem{Malleret_1959-1963} Malleret L (1959-1963) \textit{L’archéologie du Delta du Mékong}. 3 vols. Paris: EFEO, 134-137.

\bibitem{Mara_2022} Mara H (2022) \textit{Digital Archaeology}. it - Information Technology, 64(6) 2022, 193-194. https://doi.org/10.1515/itit-2022-0061

\bibitem{Matas_etal_2002} Matas J, Chum O, Urban M and Pajdla T (2002) \textit{Robust wide baseline stereo from maximally stable extremal regions}. In Proc. BMVC, 2002.

\bibitem{MacQueen_1967} MacQueen J (1967) \textit{Some methods for classification and analysis of multivariate observations}. In Proceedings of the fifth Berkeley symposium on mathematical statistics and probability (Vol. 1, No. 14, pp. 281-297).

\bibitem{Meila_2007} Meil\u{a} M (2007) \textit{Comparing clusterings – an information based distance}. Journal of Multivariate Analysis, 98: 873–895. MR2325412. doi:https://doi.org/10.1016/j.jmva.2006.11.013. 561, 566

\bibitem{Miksic_2013} Miksic J (2013) \textit{Singapore and the Silk Road of the Sea}. Singapore: NUS Press.

\bibitem{Miksic_and_Goh_2017} Miksic J and Goh GY (2017) \textit{Ancient Southeast Asia}. New York: Routledge.

\bibitem{Miksic_and_Lim_2003} Miksic J and Lim TS (2003) \textit{rchaeological Research on The Padang and in the St. Andrew’s Cathedral Churchyard: St. Andrew’s Cathedral Archaeological Research Project Progress Report Summary September 2003 – June 2004}. ARI Working Paper Series, p. 3.

\bibitem{Mitchiner_1998} Mitchiner M (1998) \textit{The History and Coinage of Southeast Asia Until the Fifteenth Century}. London: Hawkins Press.

\bibitem{Mitchiner_2002} Mitchiner M (2002) \textit{Ancient Trade and Early Coinage}. London: Cromwell Press.

\bibitem{Moore_2009} Moore E (2009) \textit{Archaeology of the Shan Plateau: The Bronze to Buddhist Transition}. Contemporary Buddhism, 10(1), 91-110.

\bibitem{Morgan_2021} Morgan C (2021) \textit{An Archaeology of Digital Things: Social, Political, Polemical}. Antiquity, 95(384), 1590-3.  http://dx.doi.org.libproxy1.nus.edu.sg/10.15184/aqy.2021.125

\bibitem{Morgan_2022} Morgan C (2022) \textit{Current Digital Archaeology}. Annual Review of Anthropology, 51, 213-231. DOI: https://doi-org.libproxy1.nus.edu.sg/10.1146/annurev-anthro-041320-114101

\bibitem{Muja_and_Lowe_2009} Muja M and Lowe DG \textit{Fast Approximate Nearest Neighbors with Automatic Algorithm Configuration}. International Conference on Computer Vision Theory and Applications.VISAPP, 2009.

\bibitem{Mus_1933} Mus P (1933) \textit{Cultes indiens et indigenes au Champa. L’Inde pre-aryenne et l’Asie des Moussons. La religion vedique et le brahmanisme. La synthese hindouiste. Formes actuelles des cultes chams. Les kut. Le culte cham des linga. Survivances et profondeur de l’influence indienne au Champa}. BEFEO, 33, 367–411.

\bibitem{Natarajan_etal_2023} Natarajan A, De Iorio M, Heinecke A, Mayer E and Glenn S (2023) \textit{Cohesion and repulsion in Bayesian distance clustering}. Journal of the American Statistical Association, 1-11.

\bibitem{Nölle_etal_2003} N\"{o}lle M, Penz H, Rubik M, Mayer K, Holl\"{a}nder I and Granec R (2003) \textit{Dagobert-a new coin recognition and sorting system}. Proceedings of the 7th International Conference on Digital Image Computing-Techniques and Applications (DICTA’03), Syndney, Australia.

\bibitem{Onwimol_2018} Onwimol W (2018) \textit{Coinage in Thailand During 4th-11th Century AD}. A Thesis Submitted in Partial Fulfilment of the Requirements for Master of Arts (Archaeology), Department of Archaeology Graduate School, Silpakorn University Academic Year 2018 Copyright of Graduate School, Silpakorn University.

\bibitem{Pamphlet_2003} Pamphlet 2003 \textit{St. Andrew’s Cathedral’s Quiet Places Project. A project to construct an extension to the existing church building.} Singapore: The QPP Building Committee, St Andrew’s Cathedral.

\bibitem{Papaioannou_etal_2002} Papaioannou G, Karabassi E and Theoharis T (2002) \textit{Reconstruction of three-dimensional objects through matching of their parts}. IEEE transactions on pattern analysis and machine intelligence, 24(1), 114-124.

\bibitem{Pizer_etal_1987} Pizer SM, Amburn P, Austin JD, Cromartie R, Geselowitz A, Greer T, ter Haar Romeny B, Zimmerman JB and Zuiderveld K (1987) \textit{Adaptive histogram equalization and its
variations}. In: Computer Vision, Graphics, and Image Processing 39.3, pp. 355–368.

\bibitem{Rand_1971} Rand WM (1971) \textit{Objective Criteria for the Evaluation of Clustering Methods}, Journal of the American Statistical Association, 66, 846–850.

\bibitem{Rasheed_and_Nordin_2015} Rasheed NA and Nordin MJ (2014) \textit{A Polynomial Function in the Automatic Reconstruction of Fragmented Objects}. Journal of Computer Science, 10(11), pp.2339-2348.

\bibitem{Rasheed_and_Nordin_2015} Rasheed NA and Nordin MJ (2015) \textit{A Survey of Computer Methods in Reconstruction of 3D Archaeological Pottery Analysis}. International Journal of Advanced Research, 3(3), 712-724.

\bibitem{Richter_etal_2011} Richter, F, CX Ries, and R Lienhart (2011) \textit{A graph algorithmic framework for the assembly of shredded documents.} Proceedings of
the 2011 IEEE International Conference on Multimedia and Expo, Barcelona, Spain, 11–15 July 2011, 1–6.

\bibitem{Rockhill_1914} Rockhill, W. W. (1914).\textit{Notes on the relations and trade of China with the eastern archipelago and the coast of the Indian Ocean during the fourteenth century.} T'oung Pao, 15, Part I, 419–47.

\bibitem{Rudin_etal_1992} Rudin LI, Stanley O and Emad F (1992) \textit{Nonlinear total
variation based noise removal algorithms}. In: Physica D: Nonlinear Phenomena 60.1, pp. 259–268.

\bibitem{Sanderson_etal_2007} Sanderson DCW, Bishop P, Stark MT, Alexander S and Penny D (2007) \textit{Luminescence dating of canal sediments from Angkor Borei, Mekong Delta, Southern Cambodia}. Quarternary Geochronology, 2 (1-4), 322-329.

\bibitem{SAA} Society of American Archaeologists (Accessed 12-05-2023). \textit{What is Archaeology}. Retrieved from: https://www.saa.org/about-archaeology/what-is-archaeology

\bibitem{Sasi_and_Sreekumar_2015} Sasi A and Sreekumar K (2015) \textit{Image based Coin Recognition System-A Survey}. International Journal of Computer Applications, 131(11).

\bibitem{Schaps_2015} Schaps D (2015) \textit{The Invention of Coinage and the Monetization of Ancient Greece}. Ann Arbor, MI: University of Michigan Press.

\bibitem{Schriebman_etal_2004} Schriebman S, Siemens R, Unsworth J (2004) \textit{A Companion to Digital Humanities}. Malden, MA: Blackwell Publishing.

\bibitem{Simonyan_and_Zisserman} Simonyan K and Zisserman A (2014) \textit{Very deep convolutional networks for large-scale image recognition}. arXiv preprint arXiv:1409.1556.

\bibitem{Stark_2004} Stark MT (2004) \textit{Pre-Angkorian and Angkorian Cambodia}. In: Southeast Asia: From Prehistory to History, Bellwood, P. and I. Glover, eds. Oxford: Routledge \& Curzon, 89-119.

\bibitem{Sutherland_and_Carson_1926-2019} Sutherland CHV and Carson RAG (ed.) (1926-2019) \textit{Roman Imperial Coinage}, 10 vols. London: Spink and Sons.

\bibitem{Taylor_2020} Taylor ZM (2020) \textit{The Computer-Aided Die Study (CADS): A Tool for Conducting Numismatic Die Studies with Computer Vision and Hierarchical Clustering}. Computer Science Honors Theses. 54. https://digitalcommons.trinity.edu/compsci\_honors/54 

\bibitem{Tin_and_Luce_1976} Tin PM, and GH Luce (1976) \textit{The Glass Palace Chronicle of the Kings of Burma}. New York: AMS Press.

\bibitem{Tuno_etal_2022} Tuno N, Mulahusi\'{c} A, Topoljak J and $\Dbar$idelija M (2022) \textit{Evaluation of handheld scanner for digitization of cartographic heritage}. Journal of Cultural Heritage, 54, 31–43.

\bibitem{Van_der_Maaten_2008} Van der Maaten L and Hinton G (2008) \textit{Visualizing data using t-SNE}. Journal of machine learning research, 9(11).

\bibitem{Vickery_1998} Vickery M (1998) \textit{Society, economics, and politics in pre-Angkor Cambodia: the 7th-8th centuries}. Tokyo: Centre for East Asian Cultural Studies for Unesco, The Toyo Bunko.

\bibitem{Vickery_2003} Vickery M (2003) \textit{Funan Reviewed: Deconstructing the Ancients.} BEFEO, 90-91, 101-143.

\bibitem{Wang_etal_2004} Wang Z, Bovik AC, Sheikh HR, Simoncelli EP (2004) \textit{Image quality assessment: from error visibility to structural similarity}. IEEE Transactions on Image Processing. 13 (4): 600–612.

\bibitem{Wicks_1985} Wicks R (1985) \textit{The Ancient Coinage of Mainland Southeast Asia.} Journal of Southeast Asian Studies, 16(2), 195-225.

\bibitem{Wicks_1992} Wicks R (1992) \textit{Chapter 4: Money and Society in Ancient Burma: Mon, Pyu, and Pagan}. In: Money, Markets and Trade in Early Southeast Asia: The Development of Indigenous Monetary Systems to AD 1400. Ithaca, NY: Cornell University Press, 111-156.

\bibitem{www.seanda.omeka.net} Wicks R [Online] \textit{Southeast Asian Numismatics Digital Archive (SEANDA)}. URL: https://seanda.omeka.net/ 

\bibitem{Zaharieva_etal_2007} Zaharieva, M, M Kampel, and S Zambanini (2007, August). \textit{Image based recognition of ancient coins.} International Conference on Computer Analysis of Images and Patterns (pp. 547-554). Springer, Berlin, Heidelberg.


\end{thebibliography}
\end{document}